\documentclass[prl,twocolumn,nofootinbib,longbibliography]{revtex4-1}
\usepackage{graphicx}
\usepackage{color}
\usepackage{epsfig}
\usepackage{amsmath}
\usepackage[caption=false]{subfig}

\begin{document}

\title{Signature of Dynamical Heterogeneity in Spatial Correlations of Particle 
Displacement  and its Temporal Evolution in Supercooled Liquids}
\author{Indrajit Tah$^{1}$}
\email{indrajittah@tifrh.res.in}
\author{Smarajit Karmakar$^{1}$}
\email{smarajit@tifrh.res.in}
\affiliation{$^1$ Tata Institute of Fundamental Research, 36/P, 
Gopanpally Village, Serilingampally Mandal,Ranga Reddy District, 
Hyderabad, 500107, India}

\begin{abstract}
The existence of heterogeneity in the dynamics of supercooled
liquids is believed to be one of the hallmarks of the glass transition. 
Intense research has been carried out in the past
to understand the origin of this heterogeneity in dynamics and a possible
length scale associated with it. We have done extensive molecular 
dynamics simulations 
of few model glass-forming liquids in three dimensions  
to understand the temporal evolution of the dynamic 
heterogeneity and the heterogeneity length scale. We find that 
although the strength of the dynamic heterogeneity is maximum at a timescale
close to characteristic $\alpha$-relaxation time of the system, 
dynamic heterogeneity itself is well-developed at timescale as short
as $\beta$-relaxation time and survives up to a timescale as long as
few tens of $\alpha$-relaxation time. Moreover, we discovered that 
temperature dependence of heterogeneity length remains the same in 
the whole time window although its absolute value changes over time
in a non-monotonic manner.
\end{abstract}

\maketitle

Dynamic heterogeneity (DH) is ubiquitous in a vast variety of natural 
processes spanning from molecular systems to biological cells and tissues.
Existence, characterization and its role in different dynamical processes
particularly in the dynamics of glass forming liquid approaching glass transition,  is an active field of research.  The dynamics of the supercooled liquid become increasingly heterogeneous \cite{RevModPhys.83.587} as the system approaches the putative glass transition point. Different regions of the system relax at time scales that differ from each other by several orders of magnitude. It is argued that the slowing down of dynamics arises mainly due to the cooperative motion of the particles. The typical size of these cooperatively rearranging regions (CRR) is believed to be of the order of few particles diameter within which motions of particles are spatially and temporarily correlated. It is already shown 
both in experiments \cite{Weeks627,annurev.physchem.51.1.99,
PhysRevLett.116.068305,doi:10.1021/ma00128a036} and in computer 
simulations \cite{DHGCG,J.Phys.Soc.Jpn,Keys2007,PhysRevLett.93.135701} 
that the mobile particles are non uniformly distributed in the system and tend to form a cluster. This inclination of clustering, as well as the size of the cluster increases as glass transition, is approached. 

Extensive studies 
\cite{RevModPhys.83.587,arcmp,annurev.physchem.51.1.99,05Berthier} have been performed in past to understand the behaviour of DH at 
the characteristic long relaxation time scale or the $\alpha$-relaxation time scale, $\tau_\alpha$
(defined later)  \cite{PhysRevE.95.022607} and only a handful studies are done at shorter 
$\beta$-relaxation time scale \cite{PhysRevLett.116.085701}. In 
\cite{PhysRevLett.116.085701}, it has been shown that $\beta$-relaxation 
time, $\tau_{\beta}$ (defined later) has a strong finite size effect, which can be rationalized if one assumes the existence of a growing correlation length. It was surprisingly found that the temperature dependence of this growing correlation length at $\beta$-relaxation time is the same as that of the heterogeneity length scale obtained at $\alpha$-relaxation time. This observation is very surprising as these time scales can differ by many orders of magnitude, especially at low temperatures.  

The main goals of this work are two-fold. The first goal is to find 
signatures of DH in the dynamics at the short time scale of the 
order of $\tau_\beta$. Then we would like to understand the subsequent 
growth and temporal evolution of DH at timescales ranging from 
$\tau_\beta$ to an order of magnitude larger than $\tau_\alpha$.  
As most of the 
research works have focused on the characterization of DH
in the $\alpha$-relaxation timescale, it is very important to 
comprehend the time evolution of DH in the intermediate as well as long-time 
scale compare to $\tau_\alpha$, in order to understand the role of
DH in glass transition. The main results of this work are the 
observation of signature of DH in the displacement 
fields of particles over $\beta$ time scale and the survival of DH  
at timescales that are larger than $\tau_\alpha$ 
by a factor of $30-40$ in the studied temperature range. We have 
also discovered that temperature dependence of the heterogeneity 
length scale remains the same throughout the studied time window, 
but the region of heterogeneity or the spatial extent of heterogeneity 
changes with time in a non-monotonic way with its maximum appearing 
at or near $\tau_\alpha$.

%In \cite{PhysRevLett.116.085701}, it has been shown that DH 
%already exists in the system at time scale as short as $\tau_{\beta}$, 

Although in \cite{PhysRevLett.116.085701}, it was shown that DH
seems to be already developed in the system at timescale close
to $\tau_\beta$, it was not immediately
clear how particle motions at this short time scale get affected
due to the presence of the heterogeneity, in other words, whether
particle motions at $\tau_\beta$ are correlated over the 
length scale of dynamic heterogeneity, is not immediately clear.
Following \cite{PhysRevLett.116.085701}, we define $\tau_\beta$ to
be the time at which logarithmic derivative of MSD develops a 
minimum which corresponds to an inflection point in the 
$\log-\log$ plot of MSD (see SI for further details).

To measure the spatial correlation and to extract the associated length scale
in the displacements of particles at $\tau_{\beta}$, 
we have implemented the procedure given in Ref. 
\cite{POOLE199851,PhysRevLett.79.2827,PhysRevLett.82.5064}. 
Note that this measure of the spatially correlated motion in super-cooled 
liquids do not depend on arbitrary cutoff parameters as already
conclusively shown in Refs.\cite{POOLE199851,PhysRevLett.79.2827,
PhysRevLett.82.5064} for DH at $\tau_\alpha$.   
The spatial correlation of the particle displacements $g_{uu}(r,\Delta t)$ 
is defined as
\begin{equation}
%\begin{split}	
%\begin{displaymath}
 g_{uu}(r,\Delta t) = \frac{\left\langle \sum\limits_{i,j=1, j\neq i}^{N}
\textbf{u}_i(t,\Delta t) \textbf{u}_j(t,\Delta t) \delta (r - |{\bf r}_{ij}(t)|)
\right\rangle}{4\pi r^2\Delta r  N \rho \langle u(\Delta t)\rangle^2},
 %\end{split}
%\end{displaymath}
\end{equation}
where $\textbf{u}_i(t,\Delta t) = \textbf{r}_i(t+\Delta t) - \textbf{r}_i(t)$
is the vector displacement of the particle between time $t$ and $ t + \Delta t$.
$\langle u(\Delta t)\rangle = \langle \frac{1}{N}\sum_{i=1}^{N} 
|\textbf{u}_i(t,\Delta t)|\rangle $, is the average displacement of 
particles within time interval $\Delta t$. 
${\bf r}_{ij}(t) = {\bf r}_{j}(t) - {\bf r}_{i}(t)$, is the distance 
between $i^{th}$ and $j^{th}$ particles. Note that our definition of 
displacement-displacement correlation is slightly different from the 
definition given in Ref.~\citep{POOLE199851}. In Ref.~\cite{POOLE199851}, scalar 
displacement of the particles are considered, whereas we have considered 
the  vector displacement of the particles \citep{JCP}. It captures the 
orientational as well as translational correlation in the particle 
displacements within the time of observation.

We have performed extensive computer simulation of four well-studied 
model glass formers in three dimensions with different inter particle 
potential over a wide range of temperatures. The model systems studied 
are the following (\romannumeral 1) 3dKA \cite{KA}, 
(\romannumeral 2) 3dR10 \cite{2dR10}, (\romannumeral 3) 3dIPL 
\cite{10PSDPRL} and (\romannumeral 4) 3dHP \cite{PhysRevLett.88.075507}. 
The details of the models and simulations are given in the Supplemental 
Information (SI). We find that the growth of DH
identified using displacement-displacement correlation function show 
strong system size dependence at least for at $\tau_\beta$. This was 
not the case in \cite{PhysRevE.79.060501,PhysRevE.84.011506} when
the correlation function was computed at $\tau_\alpha$. Thus we 
demonstrate the results of 
$g_{uu}(r,\Delta t)$ for the above models for two 
different system sizes $N = 10000$ and $N = 108000$.

\begin{figure}[htbp]
%\vskip -0.5in                                                           
%\hskip -0.2in
\includegraphics[scale=0.36]{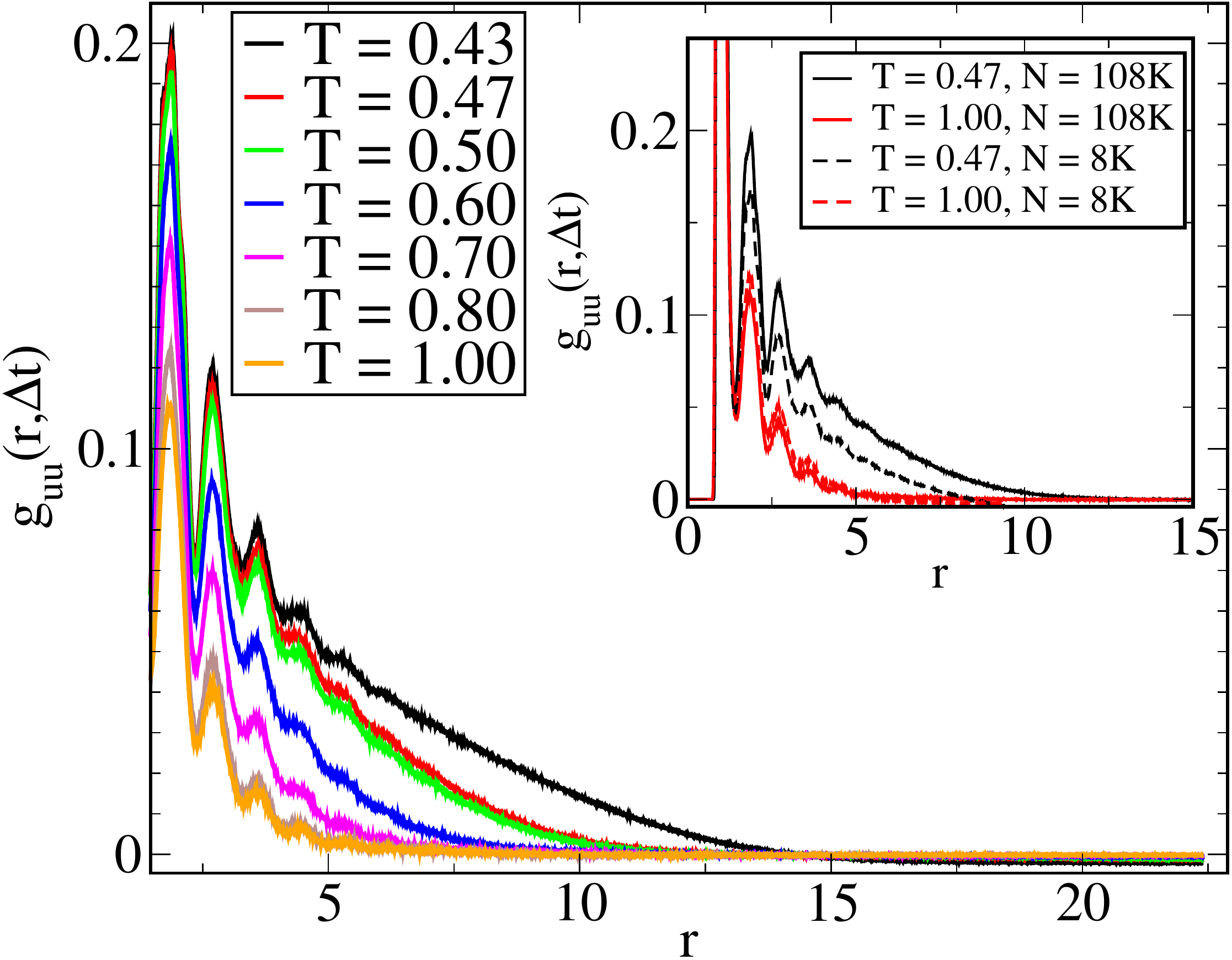}
\vskip +0.1in
\includegraphics[scale=0.41]{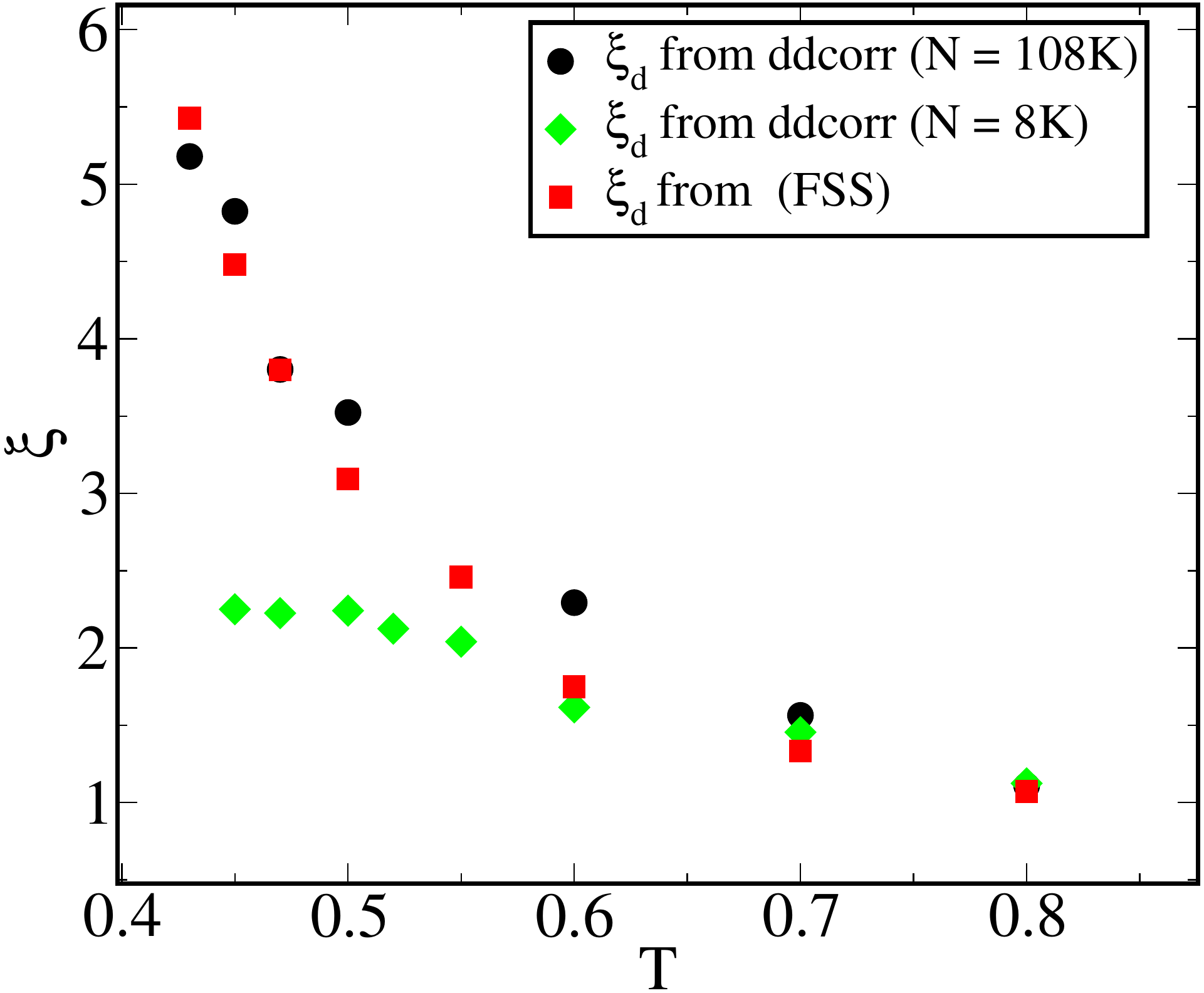}
%\vskip -0.5in
\caption{\textbf{Left Panel:} Displacement-displacement correlation 
$g_{uu}(r,\Delta t)$ at $\Delta t = \tau_\beta$ for 3dKA (N=108000). 
\textbf{Inset:} System size dependence of $g_{uu}(r,\Delta t)$ 
for the  3dKA model. 
\textbf{Right Panel:} The dynamic length scale as a function of 
time and compared with the corresponding quantities obtained using 
conventional FSS $\tau_{\beta}$.}
  \label{fig:ddcorr}
\end{figure}

In Fig.~\ref{fig:ddcorr} (left panel) we show the 
$g_{uu}(r,\Delta t)$ for $N = 108000$. It is observed that 
$g_{uu}(r,\Delta t)$ exhibits damped oscillation which is in 
agreement with previous numerical \cite{PhysRevLett.82.5064,NP1} 
as well as experimental studies \cite{0953-8984-19-20-205131}. 
The correlation function decays to zero exponentially as a function of 
distance $r$ and with decreasing temperature, the dynamics of the liquid 
becomes more heterogeneous  as the correlation  between  the  particles'
displacements in space extend up to a larger distance as shown in the 
left panel of Fig.~\ref{fig:ddcorr}. It physically means that particles 
in the liquids are moving in a co-operative fashion with a monotonically 
increasing size of the co-operative region as the temperature is lowered. 
We find a strong system size dependence in $g_{uu}(r,\Delta t)$
as shown in the inset of the left panel of Fig.~\ref{fig:ddcorr}. 
We have computed the correlation for $N = 108000$ 
and $N = 8000$ for the 3dKA model as shown in the inset of 
Fig.~\ref{fig:ddcorr}
which shows that at low temperature, relative correlation 
increases with increase in system size. For the robustness 
of our results, we have calculated the correlation for the 
other two model systems. We found the results are
qualitatively similar (see SI for further details).

In the right panel of Fig.~\ref{fig:ddcorr}, we show 
a plot of heterogeneity length scale as a function of 
temperature for the 3dKA model. 
As expected we observe strong finite size effects on the temperature 
dependence of DH length obtained from 
displacement-displacement correlation function. Heterogeneity length 
obtained from very large system size $(N = 108000)$ grow very 
similarly with dynamical length scale obtained from  finite size 
scale (FSS) of $\tau_{\beta}$ \cite{PhysRevLett.116.085701}. For $N = 8000$ 
system size one observes that the heterogeneity length grows mildly, thus
studies on smaller system size would have lead to a conclusion that 
DH is not present at $\beta$-relaxation time. After 
conclusively establishing the existence of DH at 
short time scale, we now focus on the temporal evolution of DH and 
the associated length scale across the whole range of timescales that 
can be accessed in simulation.

\begin{figure}[!h]                                                           
\hskip -0.20in
\includegraphics[scale=0.27]{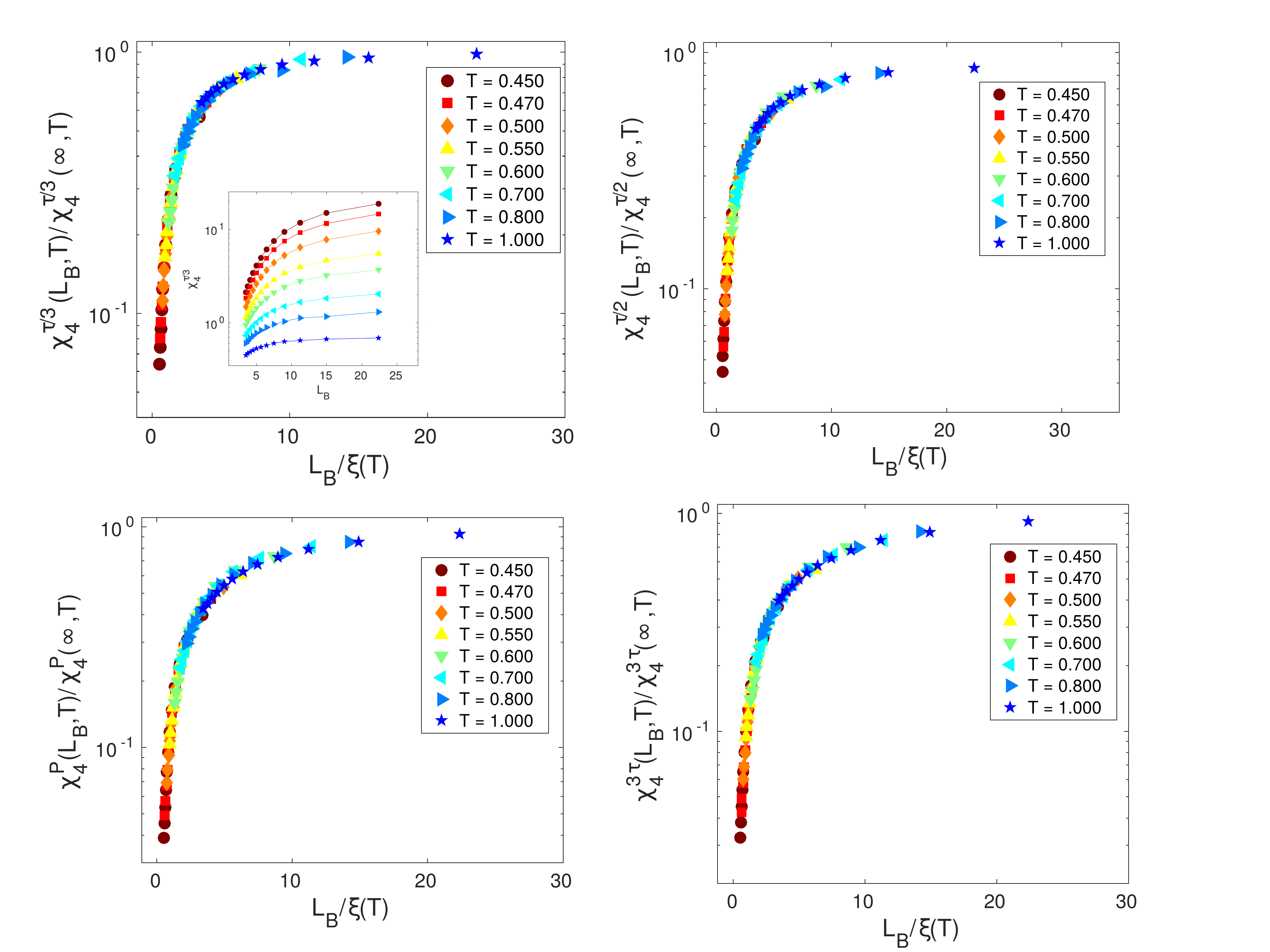}                                                      
\caption{ Block size dependence of $\chi_4$ at different time interval and collapse of data is done by rescaling the x-axis to get the length scale $\xi(T)$. Figures are shown at different time interval $\tau_\alpha/3$, $\tau_\alpha/2$, time at which intensity of heterogeneity  is maximum (close to time time $\tau_{\alpha}$) and $3\tau_\alpha$}
  \label{fig:scaling3dKA}
\end{figure}
In \cite{PhysRevE.84.011506}, although DH was studied 
over different time scales but a systematic study on the 
temperature dependence of the DH length scale was not done. 
Equipped with the method of block analysis, introduced in 
Ref.\cite{PhysRevLett.119.205502}, a systematic study of the 
temperature dependence of the dynamical length scale across 
different time scale for different model glass-forming liquids 
became computationally feasible.   
Usually, four-point correlation function, $g_4(r,t)$ 
and the corresponding susceptibilities, $\chi_4(t)$ \cite{chandan92} are
used to study DH.$\chi_4(t)$ is related to fluctuations in two-point
function, $Q(t)$. 
%The above function measure the correlation between two relaxation 
%events (density fluctuations in this case) at two different spatial 
%location, separated by distance $r$ at time $t$. 
The Fourier transform 
of $g_4(r,t)$ is known as four-point structure factor $S_4(q,t)$ and 
related to $\chi_4(t)$ as $\lim_{q \to 0}S_4(q,t)\equiv \chi_0(t)$.
$\tau_\alpha$ is defined as $\langle Q(t=\tau_\alpha)\rangle=1/e$, 
where $\langle\ldots\rangle$ denotes ensemble averages. 
(See SI for further details and definitions).
To perform the block analysis, we equilibrate a large 
system of $N = 108000$ particles 
and measure various quantities of interest by coarse-graining 
over different block sizes, $L_B$. 
We then obtain the dynamic length scale $\xi_d$ by  
FSS analysis of $\chi_4(L_B,t)$. 
%$\chi_4(L_B,t)$ is computed from the fluctuations of $Q(L_B,t)$ as 
%$\chi_4(L_B,t) = N_B\left(\langle Q(L_B,t)^2\rangle - \langle Q(L_B,t)\rangle^2\right)$.
%$Q(L_B,t)$ is the two-point density-density correlation function computed for
%the particles that are residing in the block of size $L_B$. This is defined as
%\begin{equation}
%Q(L_B,t)=\frac{1}{N_B}\sum_{i=1}^{N_B}\frac{1}{n_j}\sum_{j=1}^{n_j}w(|r_j(0)-r_j(t)|)
%\end{equation}
%$N_B$ is the number of block of size $L_B$, $n_i$ is the number of particles 
%in the $i^{th}$ block at time t=0. $w(x)$ is unity for $x \leq 0.3$ and 
%zero otherwise. 

In previous studies \cite{PhysRevLett.119.205502,PNASUSA2009}, dependence 
of $\chi_4^P(L_B)$, the maximum intensity (peak value) of dynamical 
susceptibility on block size was studied and the dynamic heterogeneity 
correlation length $\xi_d$ has been estimated by FSS analysis using the following 
scaling form
\begin{equation}
\chi_4^P(L_B,T)=\chi_4^P(\infty,T)f(L_B/\xi_d(T)),
\label{scalingAnsatz} 
\end{equation}
where $\chi_4^P(\infty,T)$ is the $L_B \to \infty$ value of dynamical 
susceptibility at a temperature $T$. In this work we have done similar 
scaling analysis for the block size dependence of the intensity of 
dynamical heterogeneity at few particular time scales  $t = \tau_{\alpha}/3,
\tau_{\alpha}/2,3\tau_{\alpha}$ to understand 
how the characteristic length scale and the exponent associated with 
dynamical susceptibility  and the heterogeneity length scale change 
with time at different temperatures.

In Fig.~\ref{fig:scaling3dKA} we have plotted the results for the 3dKA 
model system. In top panels, we reported the block size dependence of 
$\chi_4(L_B,T)$ (inset) and the scaling collapse of $\chi_4(L_B,T)$ 
at $\tau_\alpha/3$ (left panel) and $\tau_\alpha/2$ (right panels) 
time scales respectively. A similar analysis also is shown in the bottom 
panels at time scales $\tau_{\alpha}$ (left panel) and  $3\tau_\alpha$ 
(right panel).  The scaling observed in all these four cases are 
indeed very good and the calculated length scales from FSS analysis of 
the block method is found to be in good agreement with the $\xi_d$ 
obtained from the wave vector dependence of $S_4(q,t)$  
\cite{JCP119-14} (discussed in subsequent paragraph).
In FSS, $\xi_d(T)$ is known up to a multiplicative factor which is 
the same for all temperatures. In order to fix this uncertainty, $\xi_d$ 
obtain from FSS is scaled to match with $\xi_d$ obtained from 
$S_4(q,t)$ at one temperature. 
\begin{figure*}[!htpb]
\includegraphics[scale=0.40]{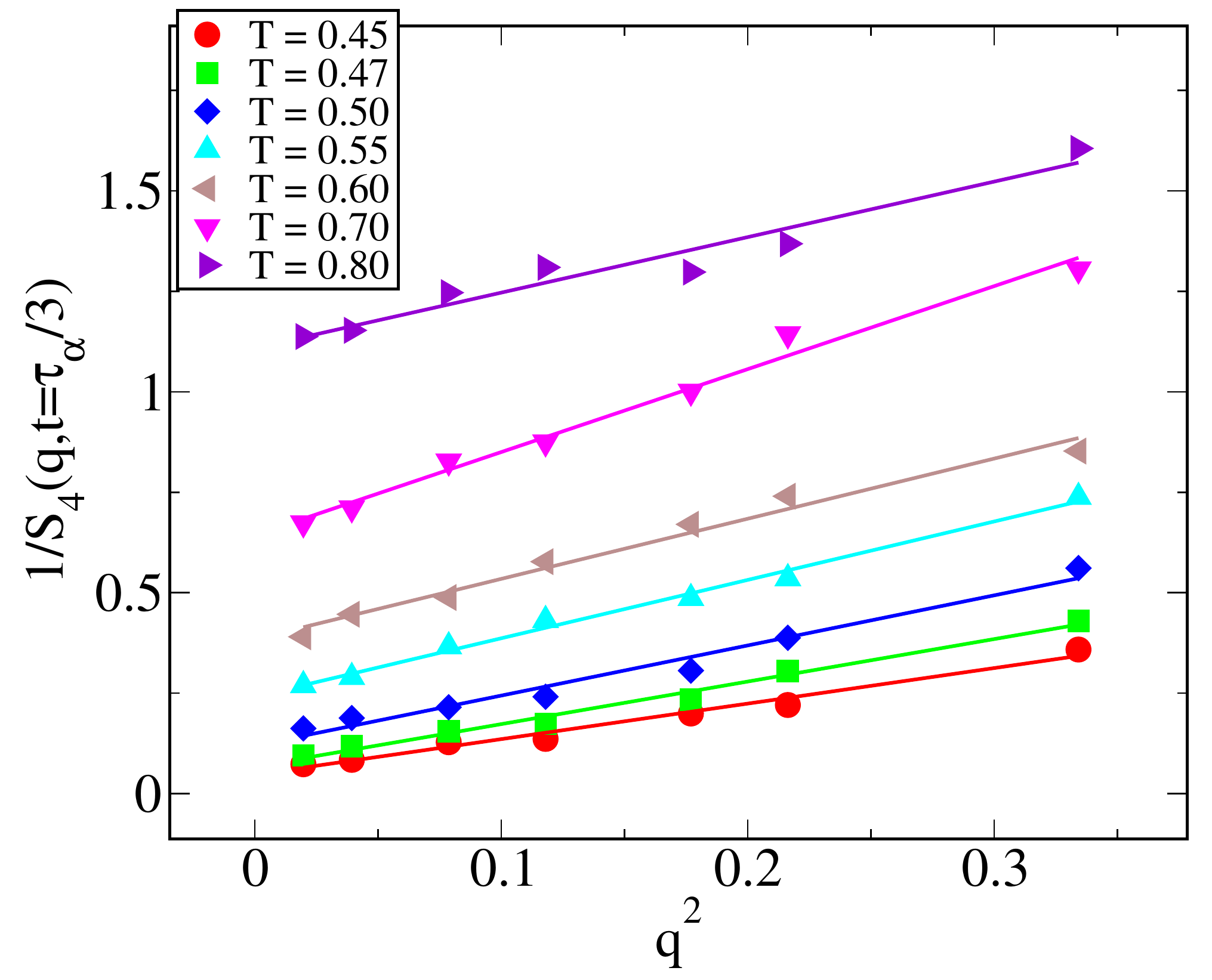}
\includegraphics[scale=0.40]{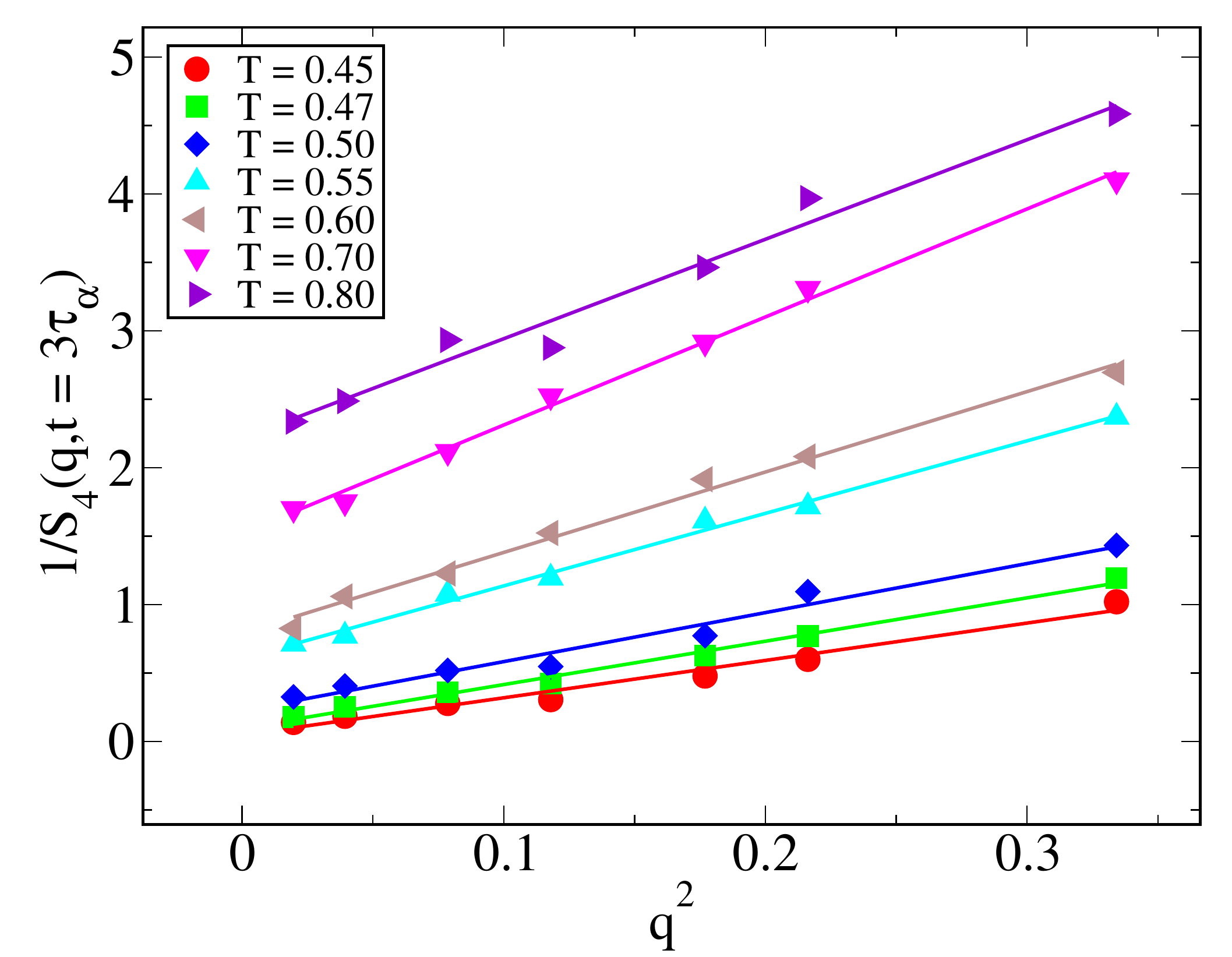}
\includegraphics[scale=0.39]{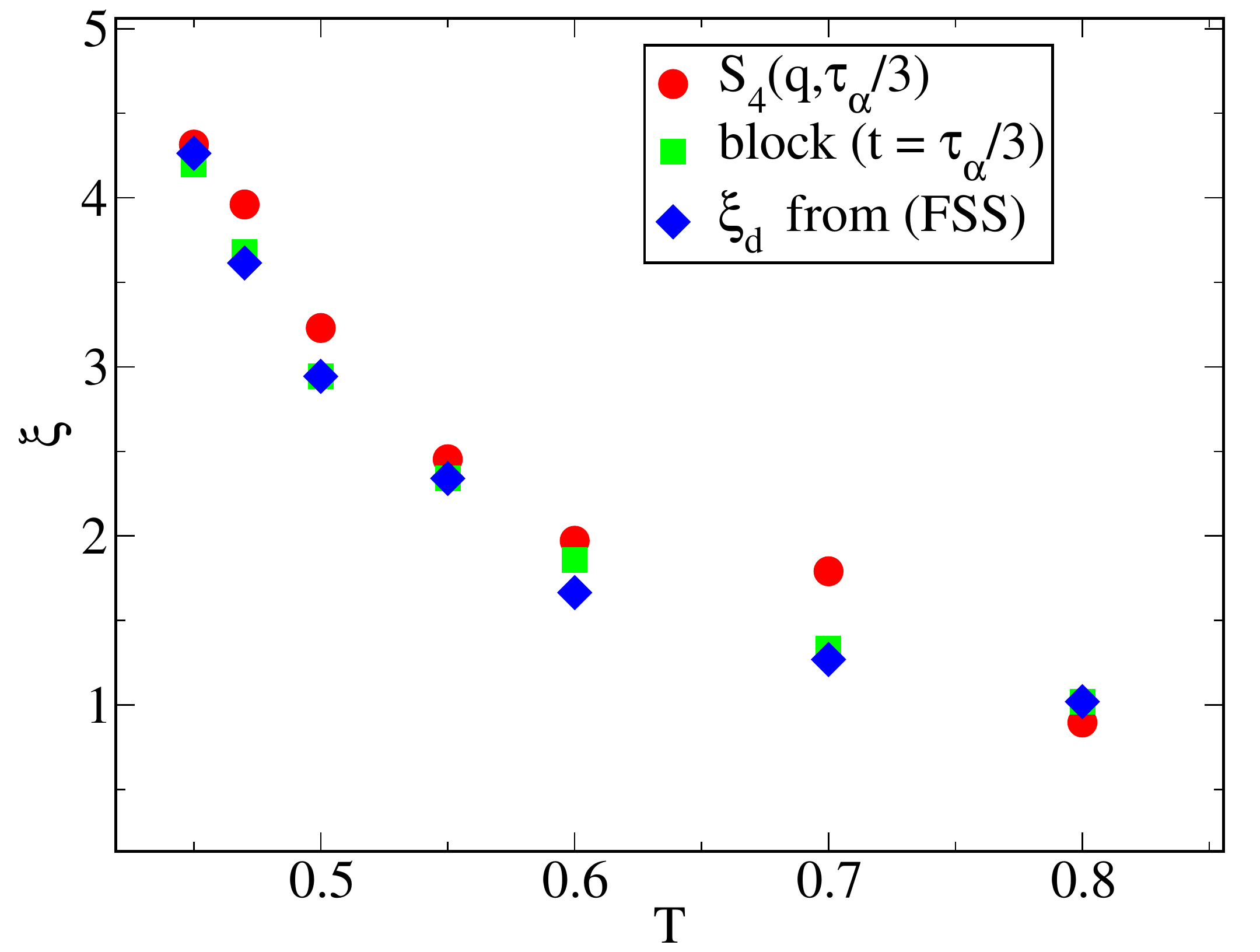}
\includegraphics[scale=0.39]{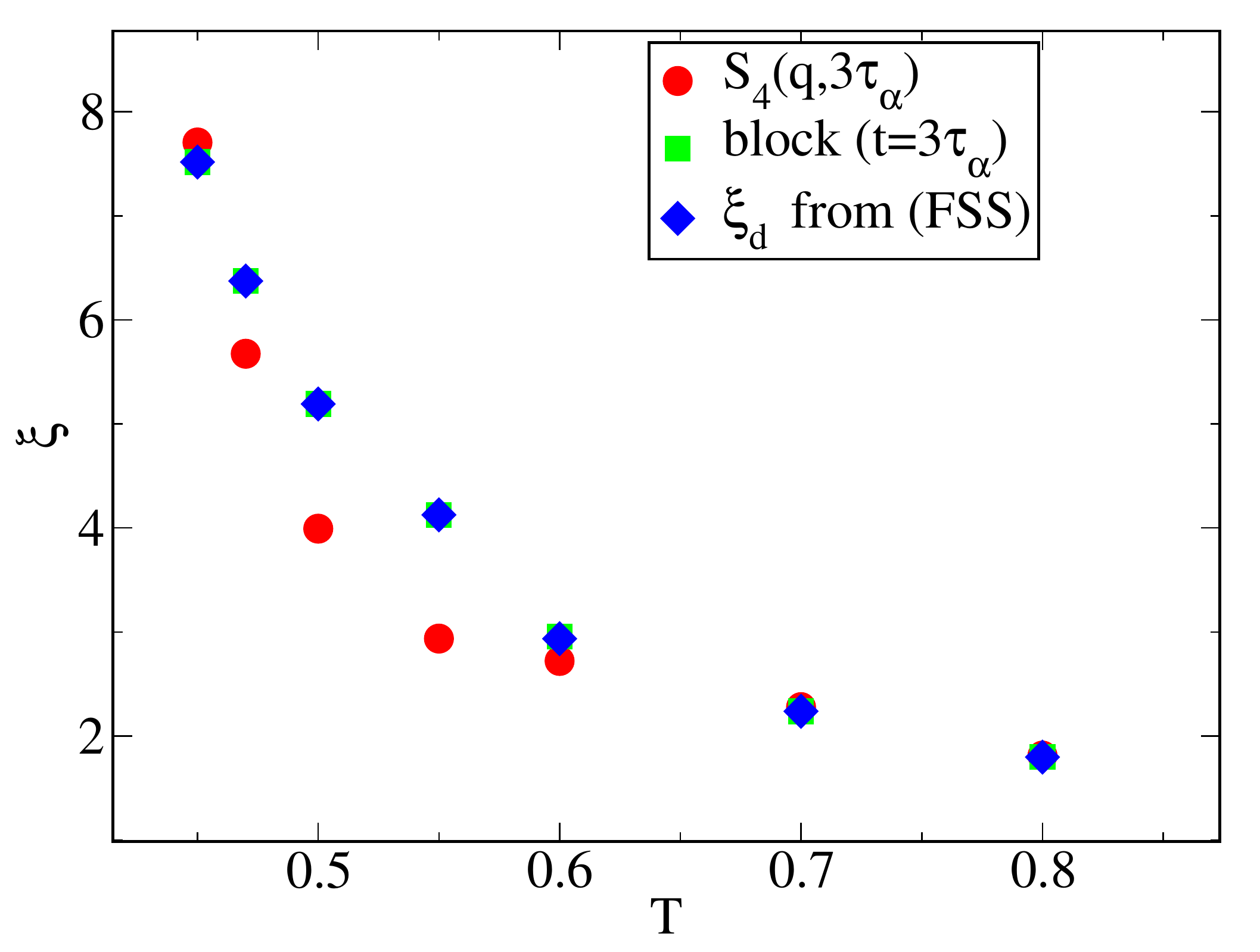}
\caption{\textbf{Top Panel:} We plot $1/S_4(q,t)$ vs $q^2$ for 3dKA 
model and we get the dynamical length scale by fitting $S_4(q,t)$ to 
the above  Ornstein-Zernicke (OZ) form $S_4(q,t) = \frac{S_4(q \to 0,t)}
{1+(q\xi)^2}$ in the range $0 < q < 0.578083$.
\textbf{Top left Panel:} length scale calculated at time interval 
$t = \tau_\alpha/3$. \textbf{Top right Panel:} length scale calculated 
at time interval $t = 3\tau_\alpha$
\textbf{Bottom Panel:} Comparison of different length scale obtained 
by  using conventional finite-size-scaling (FSS) method of $\chi_4^P$ 
\cite{PNASUSA2009} and block analysis method.}
\label{fig:com}
\end{figure*}

By fitting the q dependence of $S_4(q,t)$ for small q values to the 
Ornstein-Zernike (OZ) form $S_4(q,t) \simeq \chi_0(t)/[1+(q\xi)^2]$, 
one can also obtain $\xi_d$. It has already been shown that heterogeneity 
length scale obtained from FSS of  block method is in good agreement 
with the same obtained from $S_4(q,t)$. In top panels of 
Fig.~\ref{fig:com} we plot the wave vector dependence of the inverse 
of four point structure factor $S_4(q,t)$ for the 3dKA model for two 
different times $\tau_\alpha/3$ (left), $3\tau_\alpha)$ (right). One 
can clearly see that OZ form fits the data very well, thus the extracted 
length scale will be quite accurate. In the bottom panels of the same 
figure, the temperature dependence of the length scales computed by 
different methods are compared for $t = \tau_\alpha/3$ (left) and 
$t = 3\tau_\alpha$ (right). The legend ``$\xi_d$ from FSS" refers
to the length scale obtained from finite size scaling at $t = \tau_\beta$ 
and taken from \cite{PNASUSA2009}. Note that $\xi_d$ from FSS  are 
scaled at  $T = 0.80$. It is worth highlighting that these results suggests that 
temperature dependence of $\xi_d$ is same across timescales starting 
from $\tau_\beta$ to at least $3\tau_\alpha$. To check the robustness 
of our results, length scales for other two models (3dR10 and 3dIPL) are 
also computed and the temperature dependence of the length scale are 
found to be same over the time interval $(\tau_\beta, 3\tau_\alpha)$ 
(see SI for further details). Although variation of heterogeneity
with decreasing temperature remains same over studied timescales across 
model systems, the spatial extent of the heterogeneity is observed to 
increases up to some timescale and then starts to decrease. We are not 
able to estimate the dynamical length from FSS of block methods at 
very early time ($t < \tau_\alpha/3$) and very long time 
($t > 3\tau_\alpha$) as the intensity of $\chi_4(t)$, itself becomes 
extremely small to do any computation reliably.

\begin{figure}[!h]
\vskip +0.1in
\includegraphics[scale=0.45]{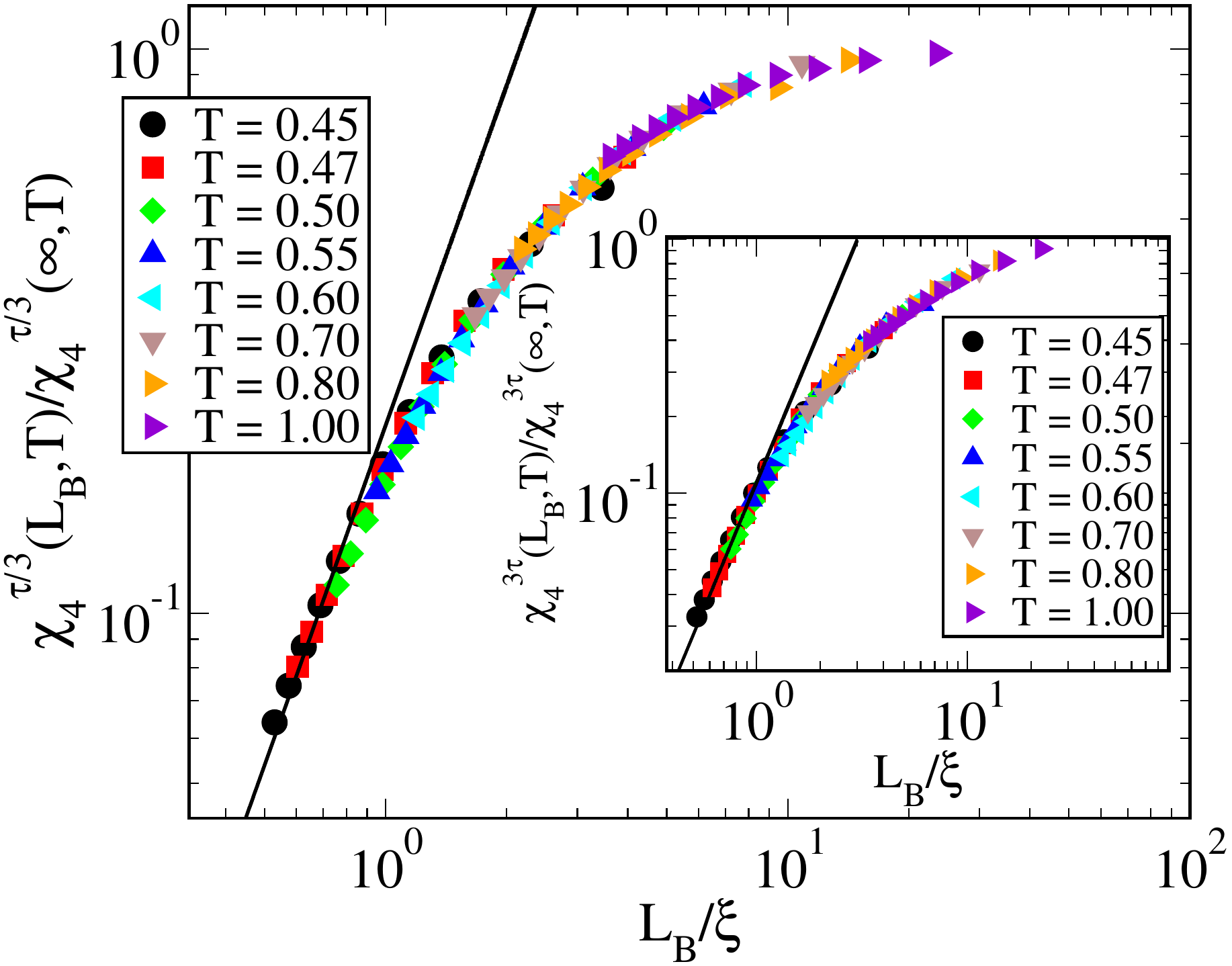}
\caption{For small $x$ scaling function $f(x) \propto x^{2-\eta}$, we find $\eta=0$ (3dKA model) at time interval $t=\tau_{\alpha}/3$ and $t=3\tau_{\alpha}$ which agrees with our previous studies.}
  \label{x4peta}
\end{figure}

Next we examine the power law relation between $\chi_4(T)$ and $\xi_d(T)$ 
across these studied timescales. According to Inhomogeneous mode coupling 
theory (IMCT) \cite{PhysRevLett.97.195701,JCPBB,JCPBB1} there exists a 
power relation between $\chi_3(\tau_{\alpha})$ (a three point correlator 
which is believed to be similar to $\chi_4$ at least in the scaling 
behaviour) and $\xi_d(\tau_{\alpha})$ as 
$\chi_3(\tau_{\alpha}) = \xi_d^{2-\eta}(\tau_{\alpha})$ where the theory predicts
exponent $2-\eta = 4$ in the $\alpha$ regime and  $2$ for the 
$\beta$ regime \cite{PhysRevLett.101.267802}. 
Following Ref.\cite{PhysRevLett.119.205502} we have done the 
scaling analysis of $\chi_4^{t^*}(T)$ to obtain the exponent $2-\eta$ 
at different times, $t^*$. 
In the large system size limit, $L_B >> \xi_d$, 
the $L_B$ dependence should disappear in the scaling relation in 
Eq.\ref{scalingAnsatz} and it should approach a constant value for $x>>1$. 
On the other hand for $\xi_d \to \infty$ and $L_B$ remains finite 
the dependence of $\chi_4$ on $\xi_d$ should go away. This implies that scaling 
function $f(x)$ should be proportional to $x^{2-\eta}$ at $x \to 0$ and 
$\chi_4^t(L_B,T)$ should grow as $L_B^{2-\eta}$. We show the results 
of such an analysis for the 3dKA model in Fig.~\ref{x4peta}. 
The exponent value 
is found to be $2-\eta \simeq 2$ for both $t = \tau_\alpha/3$ as well as at 
$3\tau_\alpha$. Note this is completely different from the exponent 
($2-\eta \simeq 4$) predicted by IMCT in the $\alpha$ regime but in 
very good agreement with the prediction at $\beta$ regime. This 
observation can be rationalized if one assumes that there will be 
less activated relaxation at a short time scale as compared to $\alpha$
relaxation time scale and MCT approximation should then be reasonable. 
Thus we can expect to have a reasonable agreement with MCT predictions at 
short timescales but not so good at longer timescales. Our results 
very nicely highlights this agreement with good quality data. 
We have done similar analysis for other models systems (3dIPL and 
3dR10) and found that the exponent $2-\eta$ is very close to 2 
(see SI for details).

Next, we look at the time evolution of the DH length scale. 
In \cite{JCP119-14}, it was shown that time dependence of $\xi_d(t)$ is 
same as $\chi_4(t)$, which is in contradiction with the 
results reported in \cite{PhysRevE.71.041505}. In  
\cite{PhysRevE.71.041505}, $\xi_d$ is found to 
increase monotonically with time, in partial  agreement 
with the results reported in \cite{PhysRevE.83.051501} 
for hard sphere systems. Moreover in \cite{PhysRevE.83.051501} 
it is found that $\xi_d$ saturates to a 
plateau at a later time. On the other hand, IMCT predicts $\xi_d$ 
to remain constant in between 
$\tau_{\beta}$ and $\tau_\alpha$.  We then look at the mutual 
time evolution of $\xi_d$ and $\chi_4$ 
for all these different model systems to understand the 
apparent contradiction in results reported in the literature. 
%For this analysis, we have only employed $N = 108000$ 
%system size to remove any finite size effects
%in the results. 

\begin{figure}[!h]
\includegraphics[scale=0.37]{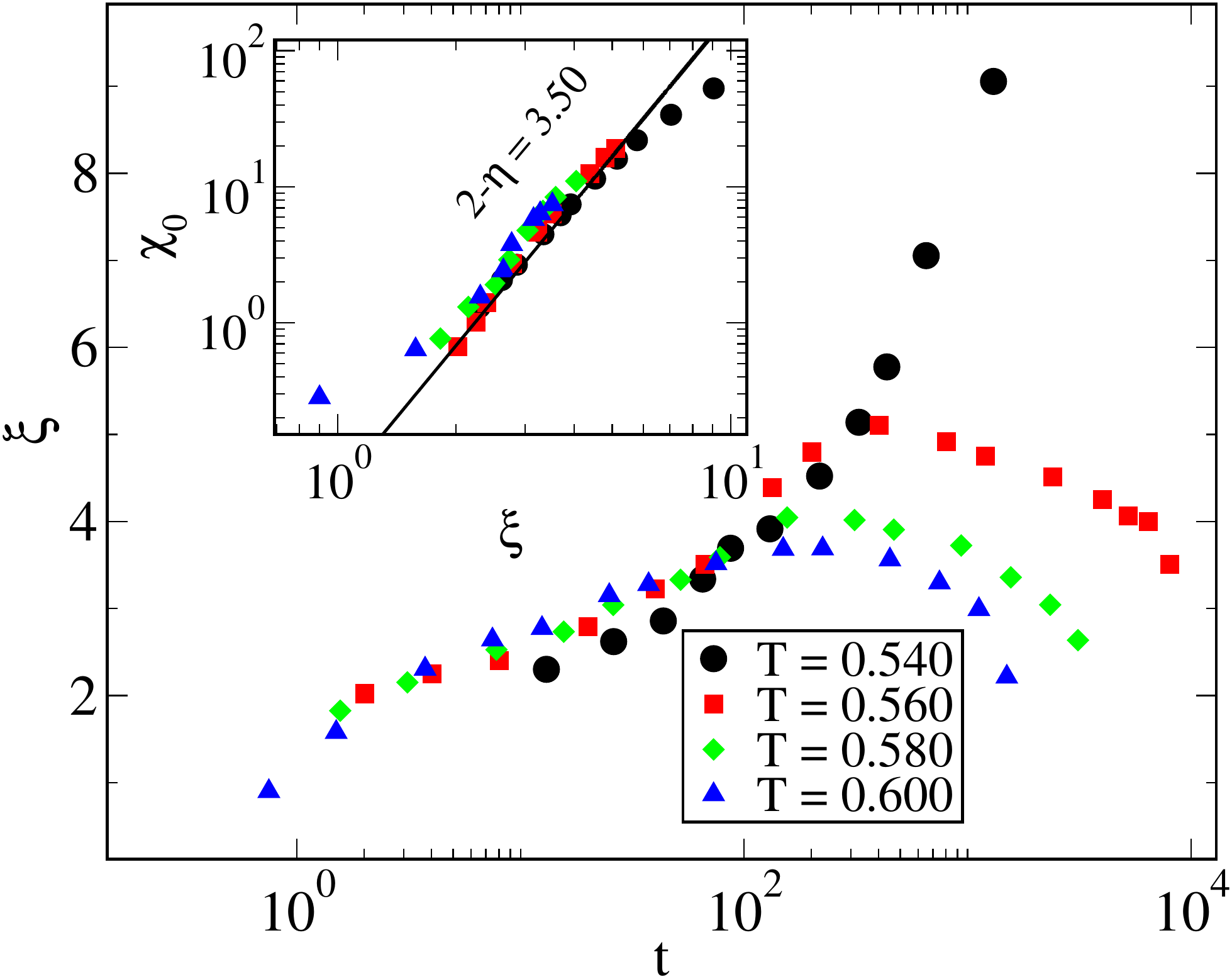}
\includegraphics[scale=0.380]{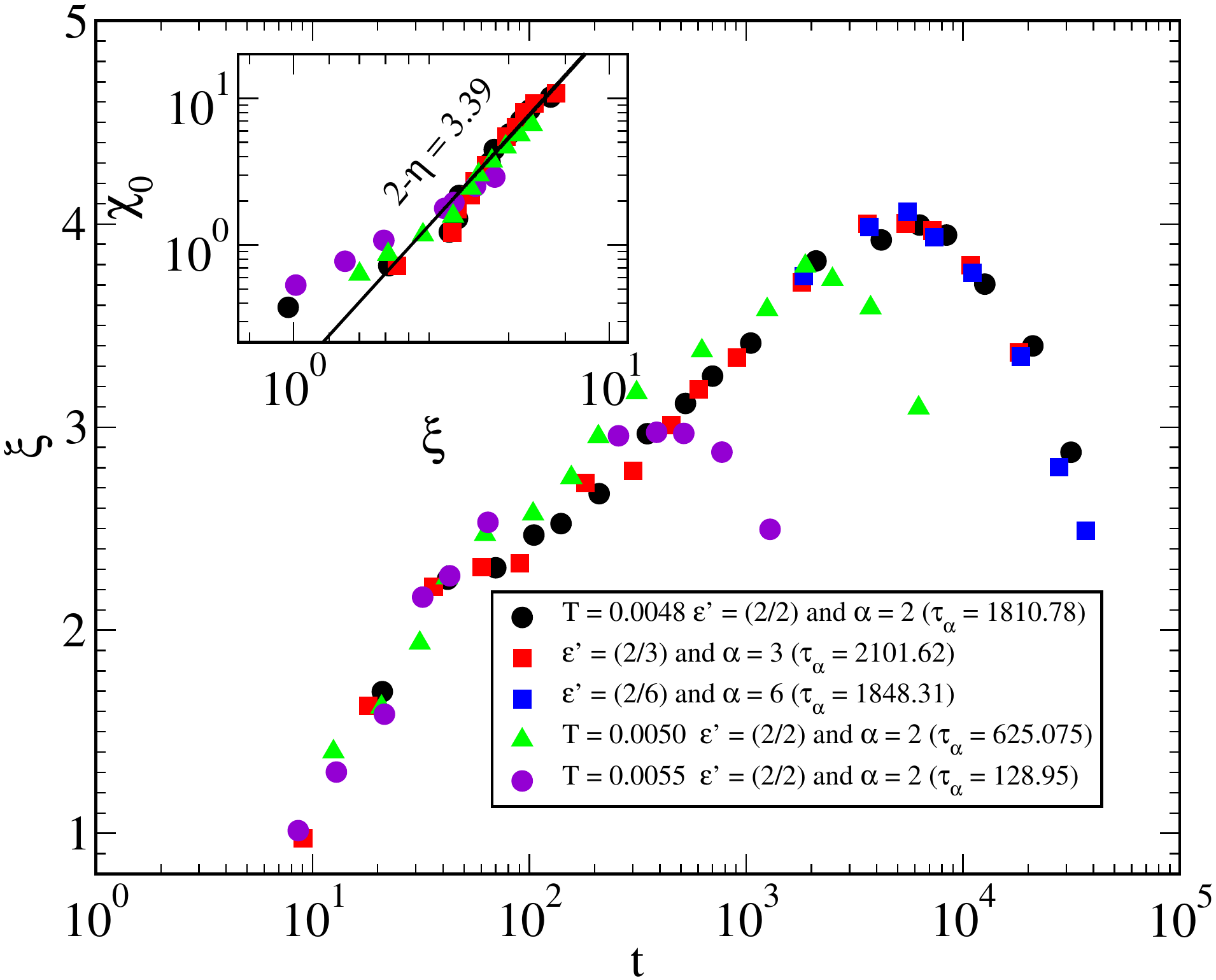}
\caption{\textbf{Top Panel:} The dynamic length scale vs time (semilog plot) for 3dR10 model. 
In the inset we show power law dependence of $\chi_4(t)$ vs $\xi(t)$. \textbf{Bottom Panel:} 
Similar plot for 3dHP model .}
%at $\epsilon = 1.0$ where $\tau_{\alpha} = 1810.78$.  
%\textcolor{red}{figure5 needs the required changes.}}
  \label{fig:difftaucollapse}
\end{figure}
In top panel of Fig.~\ref{fig:difftaucollapse}, the time 
dependence of  $\xi_d(t)$ for the 3dR10 model is shown.   
It is clear that $\xi_d(t)$ grows up to $\tau_{\alpha}$ and decreases at 
later time, in agreement with Ref. \cite{JCP119-14}.  
We then looked at the results obtained for 3dHP model (see SI)
\cite{PhysRevLett.88.075507}, which is 
one of the paradigmatic models in the context of jamming physics. 
For the 3dHP model, $\xi_d(t)$ shows
a peak at timescale close to $4\tau_\alpha$ (bottom panel of 
Fig.~\ref{fig:difftaucollapse}), which implies that $\xi_d(t)$ 
increases even though the overall strength of the heterogeneity 
decreases after $\tau_\alpha$. This suggests that a hard sphere 
like model are probably different from those models where the 
particles are treated as point particles. These observations 
although corroborate the previous observations, the reason 
for it is not immediately clear. 
A different choice of the steepness of the potential does 
change the qualitative results (see SI for detailed discussion). 
Although for 3dHP 
model $\xi_d(t) \sim \log(t)$ in agreement with
Ref.~\cite{PhysRevE.83.051501} for hard sphere systems, 
one does not see similar dependence for 3dR10 model. 
The mutual dependence between $\xi_d(t)$ and 
$\chi_4(t)$, as $\chi_4(t) \sim \xi_d(t)^{2-\eta}$, seems
to have two different regime - up to $t \sim \tau_\alpha$ 
it is power law like with exponent $2-\eta \sim 4$ 
for all the models [$4.18$ 
(3dKA), $3.50$ (3dR10) and $3.39$ for 3dHP model] but 
a extremely different one with very large exponent for
$t > \tau_\alpha$ as shown in the insets of Fig.~\ref{fig:difftaucollapse}.  
Thus it suggests that one will not be able to extract even 
approximately the dynamic heterogeneity length 
$\xi_d$ from the measurement of $\chi_4(t)$ alone as the 
exponent $2-\eta$ varies across model systems
as well as over the time window of calculation. 
%\begin{figure}[!h]
%\vskip +0.2cm
%\includegraphics[scale=0.340]{tvsxi_rescale_3dHP.pdf}
%\caption{scaling relation}
%\label{scalingAnsatz1}
%\end{figure}

Finally, to conclude, we have shown the presence of dynamic 
heterogeneity in the displacement field of particles at 
$\tau_\beta$ timescale and also highlighted the strong 
system size effect. Then 
we showed that although the absolute value of the dynamic 
heterogeneity length changes with time, the
temperature dependence of this length scale across 
timescale spanning from $\tau_\beta$ to couple of 
times $\tau_\alpha$ remain same. This is 
really surprising and may become an important test 
for validation of different theories of glass transition.
%remarkable as 
%it seems to suggest that changes in the
%dynamic heterogeneity length scale due to temporal evolution 
%or change in temperature, are 
%probably obeying the following scaling relation: 
%$\xi_d(T,t) \sim \xi^{*}(T)\mathcal{F}(t/\tau_\alpha)$.
%Fig.\ref{scalingAnsatz1}, indeed suggests the existence 
%of such a scaling function. 
We also found that absolute value of $\xi_d$ reaches its maximum value 
at $t \sim \tau_\alpha$ for 3dKA, 3dIPL 
and 3dR10 models but it does so at $ \sim 4\tau_\alpha$ 
for 3dHP models. These observations
are indeed extremely intriguing and warrants further 
research as to why heterogeneity length scale continues 
to increase at timescales where $\chi_4$ already decays 
to small values for soft sphere and hard sphere model systems. 

\bibliography{ddcorr} 
\bibliographystyle{apsrev4-1}
%bibliographystyle{h-physrev}
%\begin{thebibliography}{10}
\end{document}

\title{Signature of Dynamical Heterogeneity in Spatial Correlations of Particle 
Displacement  and its Temporal Evolution in Supercooled Liquids - Supplementary
Information}
\author{Indrajit Tah$^{1}$}
\email{indrajittah@tifrh.res.in}
\author{Smarajit Karmakar$^{1}$}
\email{smarajit@tifrh.res.in}
\affiliation{$^1$ Tata Institute of Fundamental Research, 36/P, 
Gopanpally Village, Serilingampally Mandal,Ranga Reddy District, 
Hyderabad, 500107, India}

\maketitle

\section{Models and Simulation Details}

\label{modelsAndSim}
We have studied four different model glass forming liquids in three 
dimensions. The model details are given below:
\vskip +0.3cm
\noindent{\textbf{3dKA Model:}} 

This is the well known 80:20 binary  mixture of Lennard-Jones 
particles \cite{KA}. This model was first introduced by Kob-Anderson to 
simulate $Ni_{80}P{20}$. 

\noindent The interaction potential is given by
\[ V_{\alpha\beta}(r) = 4\epsilon_{\alpha\beta}[(\frac{\sigma_{\alpha\beta}}{r})^{12} - (\frac{\sigma_{\alpha\beta}}{r})^6]\]
where $\alpha,\beta \in \{ A,B \}$ and $\epsilon_{AA} = 1.0$, $\epsilon_{AB} = 1.5$, $\epsilon_{BB} = 0.5$; 
$\sigma_{AA} = 1.0$, $\sigma_{AB} = 0.80$,   $\sigma_{BB} = 0.88$. the number density is $\rho = 1.20$. The interaction potential was cut off at $2.50\sigma_{\alpha\beta}$. We use a quadratic polynomial to 
make the potential and its first two derivatives smooth at the cutoff distance. Length, energy and time 
scale are measured in 
units of $\sigma_{AA}, \epsilon_{AA}$ and $\sqrt{\frac{\sigma_{AA}^2}{\epsilon_{AA}}}$. 
%For Argon these units 
%corresponds to a length of $3.4 \AA$, an energy of $120Kk_{\beta}$ and time of  $3\times 10^{-13} s$. 

\vskip +0.5cm
\noindent{\textbf{3dR10 Model:}} 

This is a \textbf{50:50} binary mixture \cite{R10} 
with repulsive interaction potential, defined as 
\[ V_{\alpha\beta}(r) = \epsilon_{\alpha\beta}[(\frac{\sigma_{\alpha\beta}}{r})^{10}]\]
Here $\epsilon_{\alpha\beta} = 1.0$, $\sigma_{AA} = 1.0$, $\sigma_{AB} = 1.18$,   
$\sigma_{BB} = 1.40$. The number density is $\rho = 0.81$. The interaction potential is cut-off at 
$1.38\sigma_{\alpha\beta}$. 

\vskip +0.5cm
\noindent{\textbf{3dIPL Model:}} 
The model was first studied 
in \cite{10PSDPRL}. The interaction potential is given by 
\[ V_{\alpha\beta}(r) = 1.945\epsilon_{\alpha\beta}[(\frac{\sigma_{\alpha\beta}}{r})^{15.48}]\]
All the parameters and interaction cut-offs are the same as 3dKA model. Though this 
model has a purely repulsive interaction but the interaction domain is much larger than 
that in the 3dR10 model. The  interaction range plays an major role in determining both dynamic and 
mechanical properties of the system \cite{PhysRevLett.93.105502,PhysRevE.83.046106,Yan6307}.
\vskip +0.5cm

\noindent{\textbf{3dHP Model:}} 

We have simulated a model system that interpolates between 
finite-temperature glasses and hard-sphere glasses and has 
been studied extensively in the context of jamming physics. This 
is a 50:50 binary mixture with diameter ratio of $1.4$ 
\cite{PhysRevLett.88.075507}.

\noindent The interaction potential is given by 
\[ V_{\alpha\beta}(r) = \frac{\epsilon}{\alpha} \left[1 - \left(\frac{{r}}{\sigma_{\alpha\beta}}\right)\right]^{\alpha}\] for $r_{\alpha\beta} < {\sigma_{\alpha\beta}}$ and $V_{\alpha\beta}(r)$=0 otherwise, where $\sigma_{\alpha\beta} = \frac{(\sigma_{\alpha}+\sigma_{\beta})}{2}$.
We simulated this model at a constant density $\rho$ = 0.82 and temperature 
as a control parameter. The value of $\epsilon$ is chosen to be $2.0$. 
We also did simulation by varying the $\alpha$ value. The $\alpha$ value 
we have used are the following, $(\alpha=2$ for harmonic sphere$, 
\alpha = 3$ and $\alpha= 6)$. Many studies of this model have been 
done at finite temperatures \cite{0295-5075-86-1-10001,
PhysRevE.80.021502,nature} and it finds experimental realizations in 
soft colloids, emulsions and grains \cite{book1}. For this model we 
have done simulation of a very large system with $N = 108000$ particles.  

We have done NVT simulations for all the model systems using velocity-Verlet 
integration scheme. For all the simulations we have first equilibrated our systems at 
least for $200\tau_\alpha$ (defined later) and stored data for similar simulation time.  

\section{Correlation function: $Q(t)$ and $\chi_4(t)$}
The four-point dynamical susceptibility $\chi_4(t)$ \cite{chandan92} is 
defined in terms of the fluctuations of the two point overlap correlation 
function $Q(t)$ as  
\begin{equation}
\chi_4(t) = N\left[\langle Q^2(t)\rangle - \langle Q(t)\rangle^2 \right].
\end{equation}
The function $Q(t)$ is defined as
\begin{equation}
Q(t) = \frac{1}{N}\sum_{i=1}^{N}w\left(|\vec{r}_i(t) - \vec{r}_i(0)|\right),
\end{equation}
This function measures the overlap of a configuration of particles at 
a given initial time $(t=0)$ with the configuration at a later time $t$.
$w(x)$ is a window function 
defined as is $w(x) = 1$ for $x < a$ and $0$ otherwise.
The window function $w(x)$ is introduced to remove the apparent 
de-correlation that might happen due to the vibrational motion 
of particles inside the cages formed by their neighbours. The 
particular choice of $a$ is not very important, and in our studies 
we choose $a=0.3$ which corresponds to the 
plateau value of the mean square displacement.
The $\alpha$-relaxation time $\tau_\alpha$ is defined as 
$\langle Q(t=\tau_\alpha)\rangle=1/e$, where $\langle\ldots\rangle$ denotes
ensemble averages.

\section{$\beta$-relaxation time : $\tau_{\beta}$}

\begin{figure}[!h]
\begin{center}
%\centering
%\hskip -0.2cm
\includegraphics[scale=0.38]{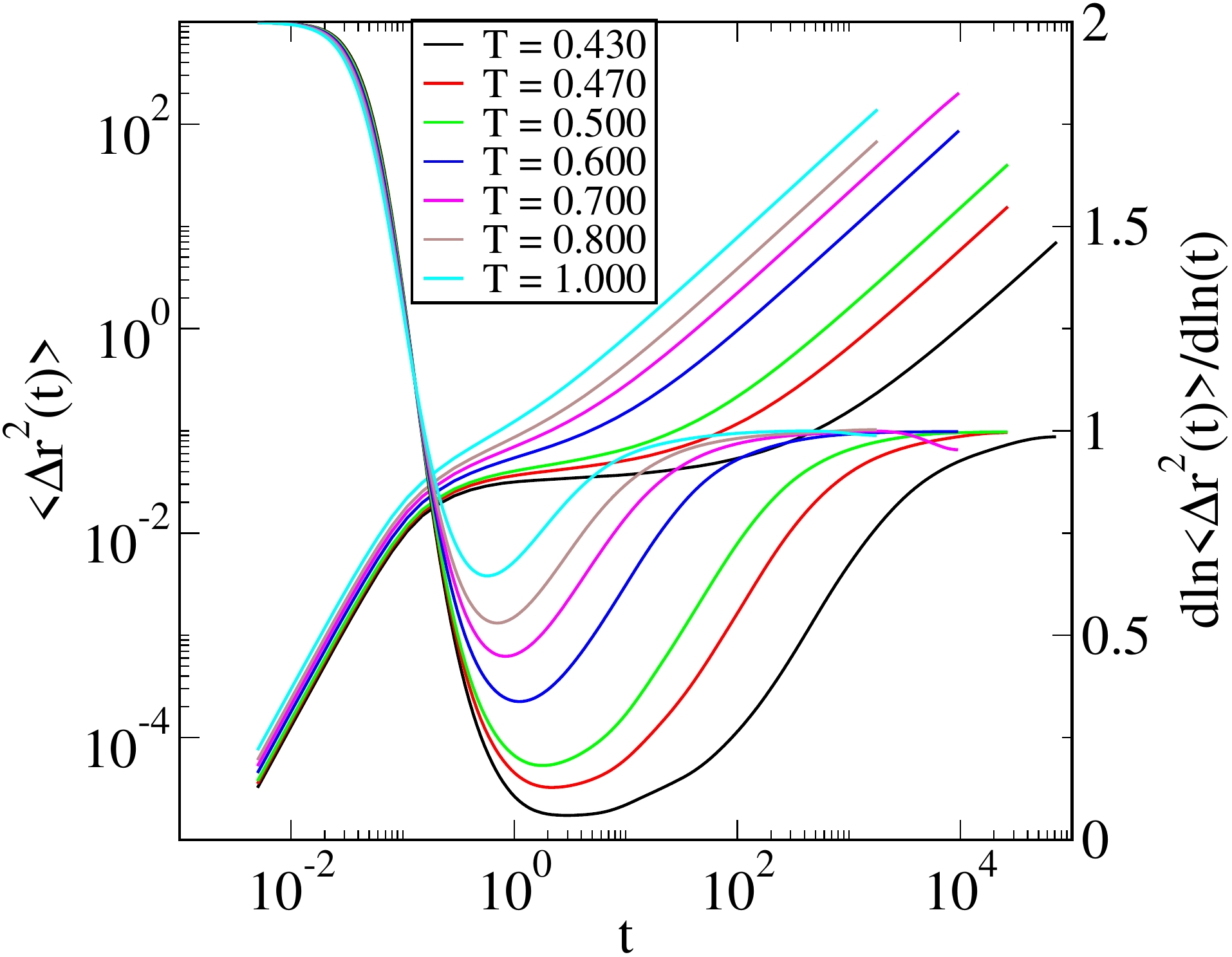}

\caption{Mean square displacement (MSD) and its logarithm derivative
with respect to logarithm of time  for 3dKA model for N = 108000. The $\tau_{\beta}$ 
time scale is calculated from the minima of the derivative.}
\label{taubeta}
\end{center}
\end{figure}
To calculate $\tau_{\beta}$, we follow the procedure of Ref.~\cite{Larini2007,PhysRevLett.116.085701}, where $\tau_{\beta}$ is calculated from 
the inflection point in the $\log$-$\log$ plot of mean squared displacement (MSD) 
$\langle r^2(t) \rangle$.  Mean square displacement $(\langle r^2(t) \rangle)$ is defined as 
\begin{equation}
\langle |\Delta r(t)|^2 \rangle = \left\langle \frac{1}{N}\sum_{i=1}^N 
|\vec{r}_i(t) - \vec{r}_i(0)|^2\right\rangle.
\end{equation}
In Fig.~\ref{taubeta}, we plot the MSD in $\log$-$\log$ as a function of time and 
log-derivative of MSD with time, $d\log{\langle |\Delta r(t)|^2 \rangle}/d\log{t}$.
From the minimum in the derivative, one can estimate the $\tau_{\beta}$ time scale without much uncertainty.

\section{Displacement-displacement correlation}
To validate our results we measure the spatial correlation of 
particle displacement and associated length scale in the 
displacements correlation of particles at $\tau_{\beta}$ time scale
for 3dIPL model system.

\begin{figure*}%[!h]
\begin{center}
\includegraphics[scale=0.27]{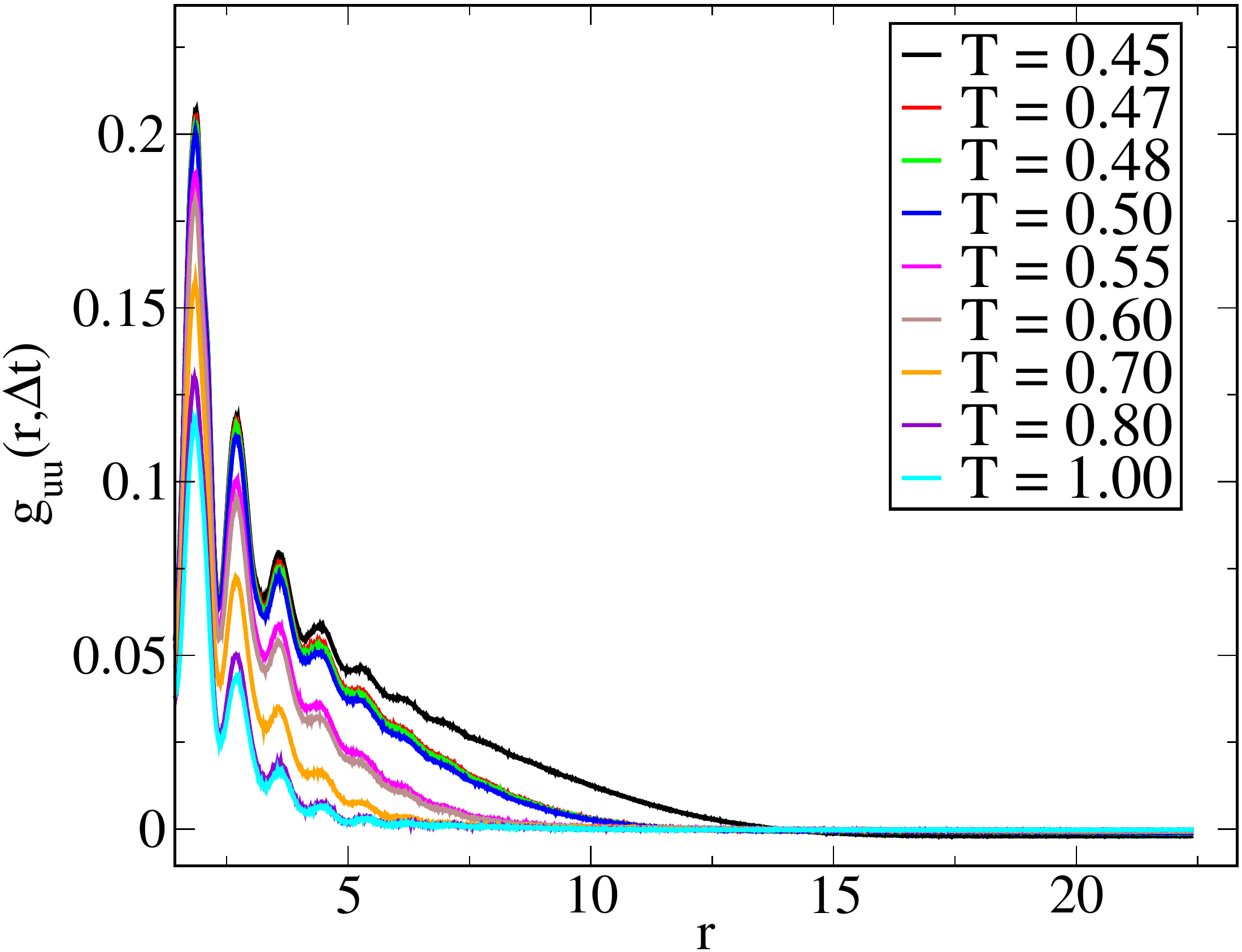}%
\hskip +0.2 cm                
\includegraphics[scale=0.25]{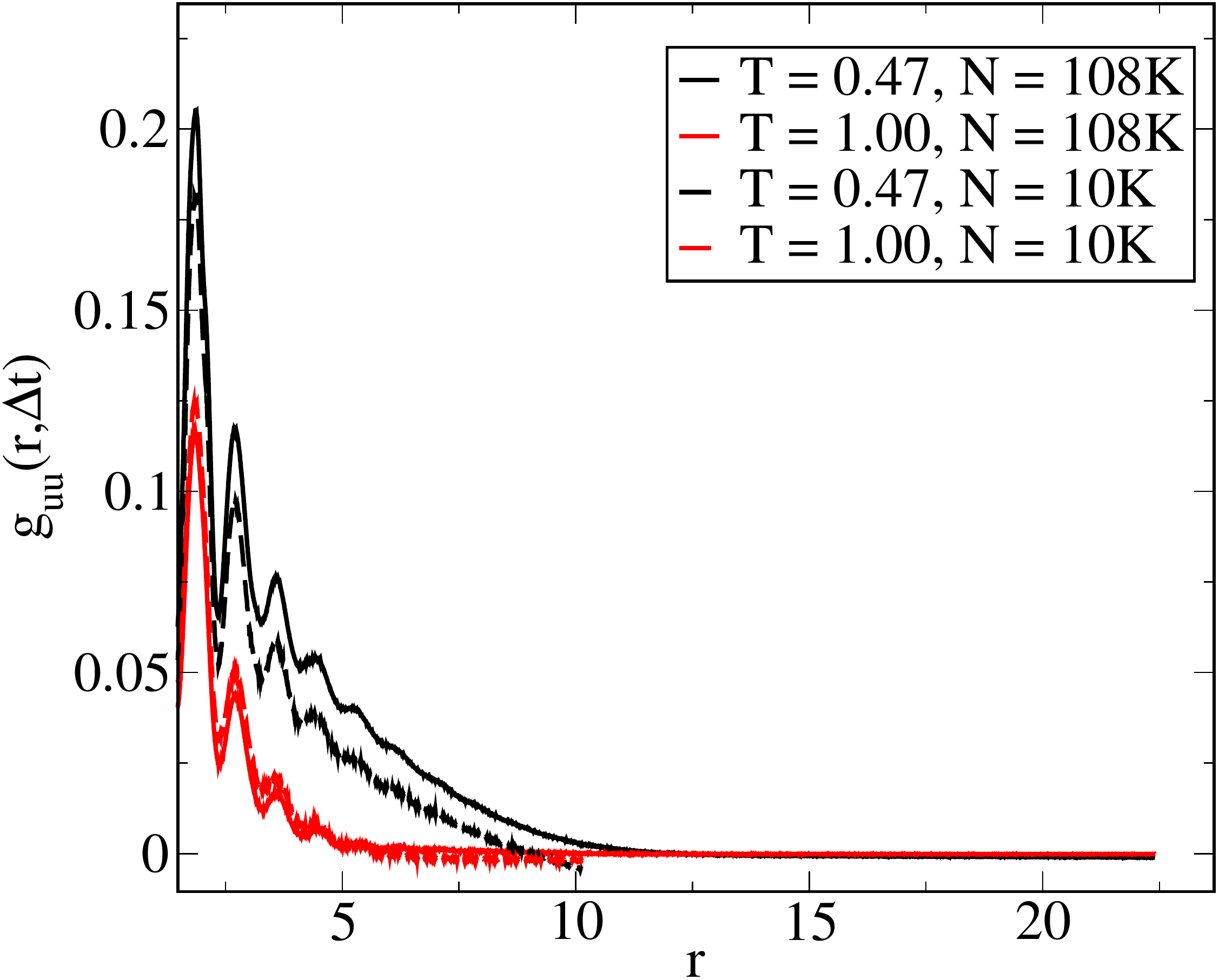}
\hskip +0.2 cm
\includegraphics[scale=0.23]{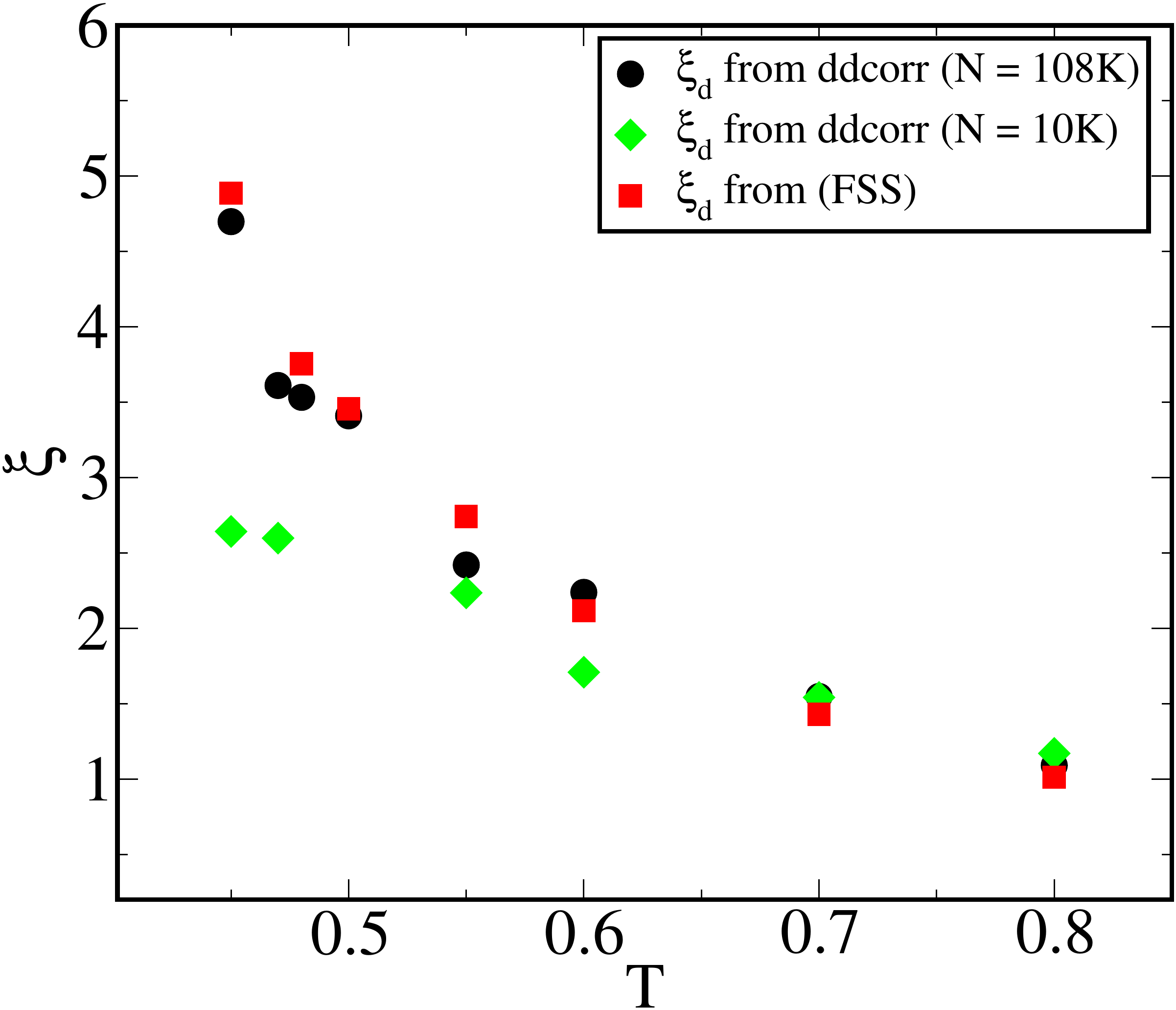}
  \caption{\textbf{Left Panel:} Displacement-displacement correlation $g_{uu}(r,\Delta t)$ at $\Delta t = \tau_\beta$ for 3dIPL model systems (N=108000). \textbf{Middle Panel:} System size dependence of $g_{uu}(r,\Delta t)$ for the  3dIPL model. \textbf{Right Panel:} The dynamic length scale as a function of temperature and compared with the corresponding quantities obtained using conventional FSS of $\tau_{\beta}$ \cite{PhysRevLett.116.085701}.}
  \label{fig:ddcorr}
\end{center}
\end{figure*} 

We find strong system size dependence of the displacement-displacement 
correlation as 
shown for two different system sizes ($N = 108000$ and $N = 10000$) 
in middle panel of 
Fig.~\ref{fig:ddcorr}. In right panel of Fig.~\ref{fig:ddcorr}, 
we plot the heterogeneity length scale as a function of temperature for
3dIPL model. Like 3dKA model, we also observe strong system size effects on the 
temperature dependence of DH length obtained from displacement-displacement 
correlation function. Heterogeneity length obtained from very large system 
size ($N = 108000$) grow 
very similarly with dynamical length scale obtained from finite size scale 
(FSS) of $\tau_{\beta}$ \cite{PhysRevLett.116.085701}.

\section{Block analysis to calculate the dynamical length scale}

To calculate the dynamical length scale, we perform the finite size scaling
(FSS) analysis of dynamical susceptibility $\chi_4(L_B,t)$ following the 
procedure of Ref.~\cite{PhysRevLett.119.205502}. $\chi_4(L_B,t)$ is computed from 
the fluctuations of $Q(L_B,t)$ as
\begin{equation}
\chi_4(L_B,t) = N_B\left(\langle Q(L_B,t)^2\rangle - \langle Q(L_B,t)\rangle^2\right).
\end{equation}
$Q(L_B,t)$ is the two-point density-density correlation function computed for
the particles that are residing in the block of size $L_B$ at the chosen 
time origin $t = 0$. 
$Q(L_B,t)$ is defined as
\begin{equation}
Q(L_B,t)=\frac{1}{N_B}\sum_{i=1}^{N_B}\frac{1}{n_j}\sum_{j=1}^{n_j}w(|r_j(0)-r_j(t)|),
\end{equation}
where $N_B$ is the number of blocks of size $L_B$, $n_i$ is the number of 
particles in the $i^{th}$ block at time $t = 0$.  

The dynamic heterogeneity correlation length $\xi_d$ has been estimated by FSS 
analysis using the following scaling form
\begin{equation}
\chi_4^t(L_B,T)=\chi_4^t(\infty,T)f(L_B/\xi_d(T)),
\label{scalingAnsatz} 
\end{equation}
where $\chi_4^t(\infty,T)$ is the $L_B \to \infty$ value of dynamical 
susceptibility at  temperature $T$ and time $t$.

\section{Temperature dependence of the dynamics heterogeneity length scale over different time for 3dR10 and 3dIPL model system}

\begin{figure*}%[!h]
\begin{center}
\includegraphics[scale=0.51]{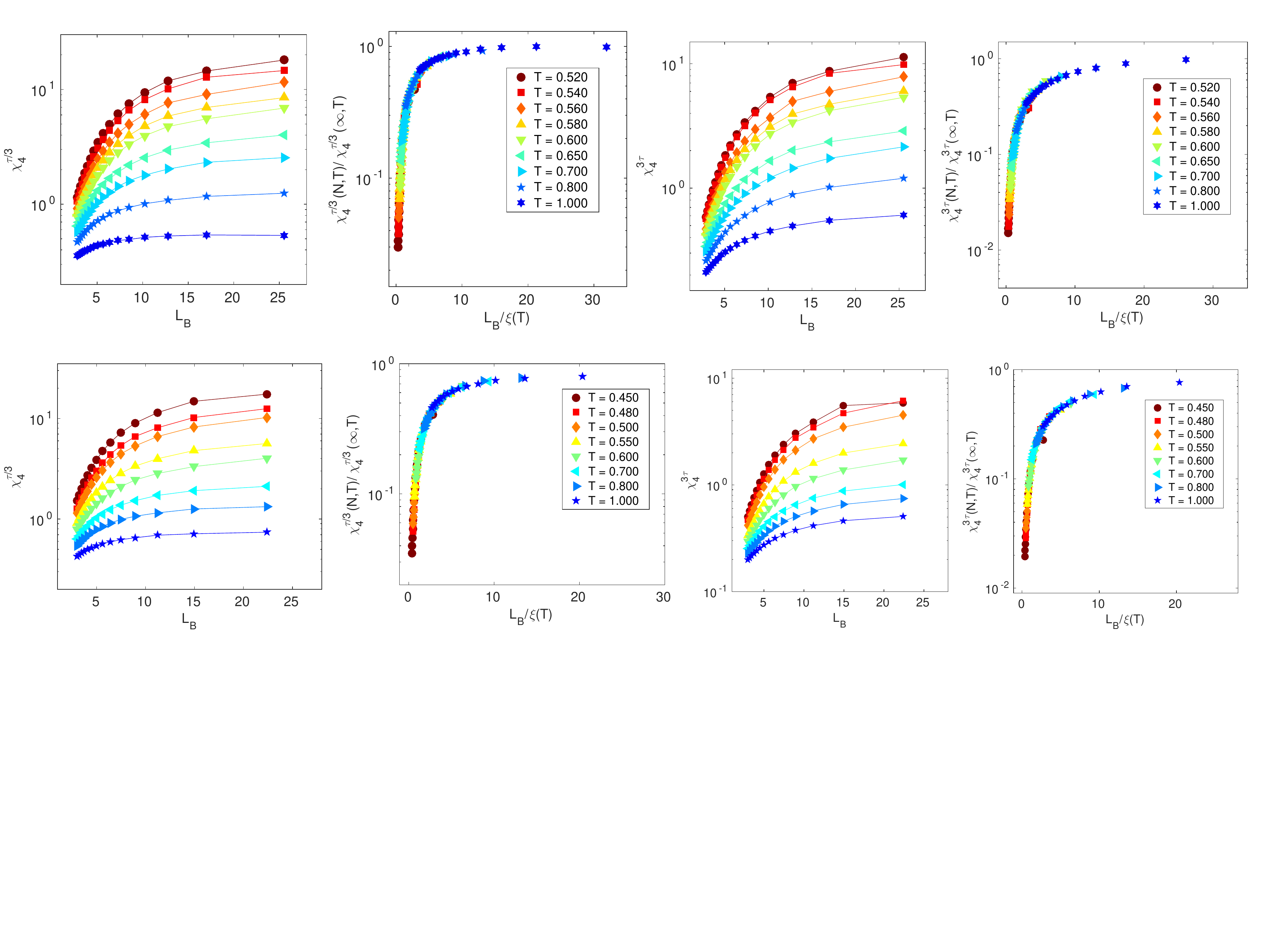}                                                           
%\includegraphics[scale=0.28]{chitauby3_3dR10.pdf}%                
%\includegraphics[scale=0.26]{chitau3_3dR10.pdf}\\
%\includegraphics[scale=0.26]{chitauby3_3dIPL.pdf}%
%\hskip -0.2in              
%\includegraphics[scale=0.24]{chitau33dIPL.pdf}  %                                                        
\vskip -1.7in
  \caption{\textbf{Top Panel:} Block size dependence of $\chi_4$ at time interval $\tau_\alpha/3$ and $3\tau_\alpha$ and its scaling collapse to get the length scale $\xi(T)$ for 3dR10 model. \textbf{Bottom panel:} Similar plot for 3dIPL model.}
  \label{fig:scaling3dR10}
\end{center}
\end{figure*}

\begin{figure*}%[!h]
\begin{center}
\includegraphics[scale=0.51]{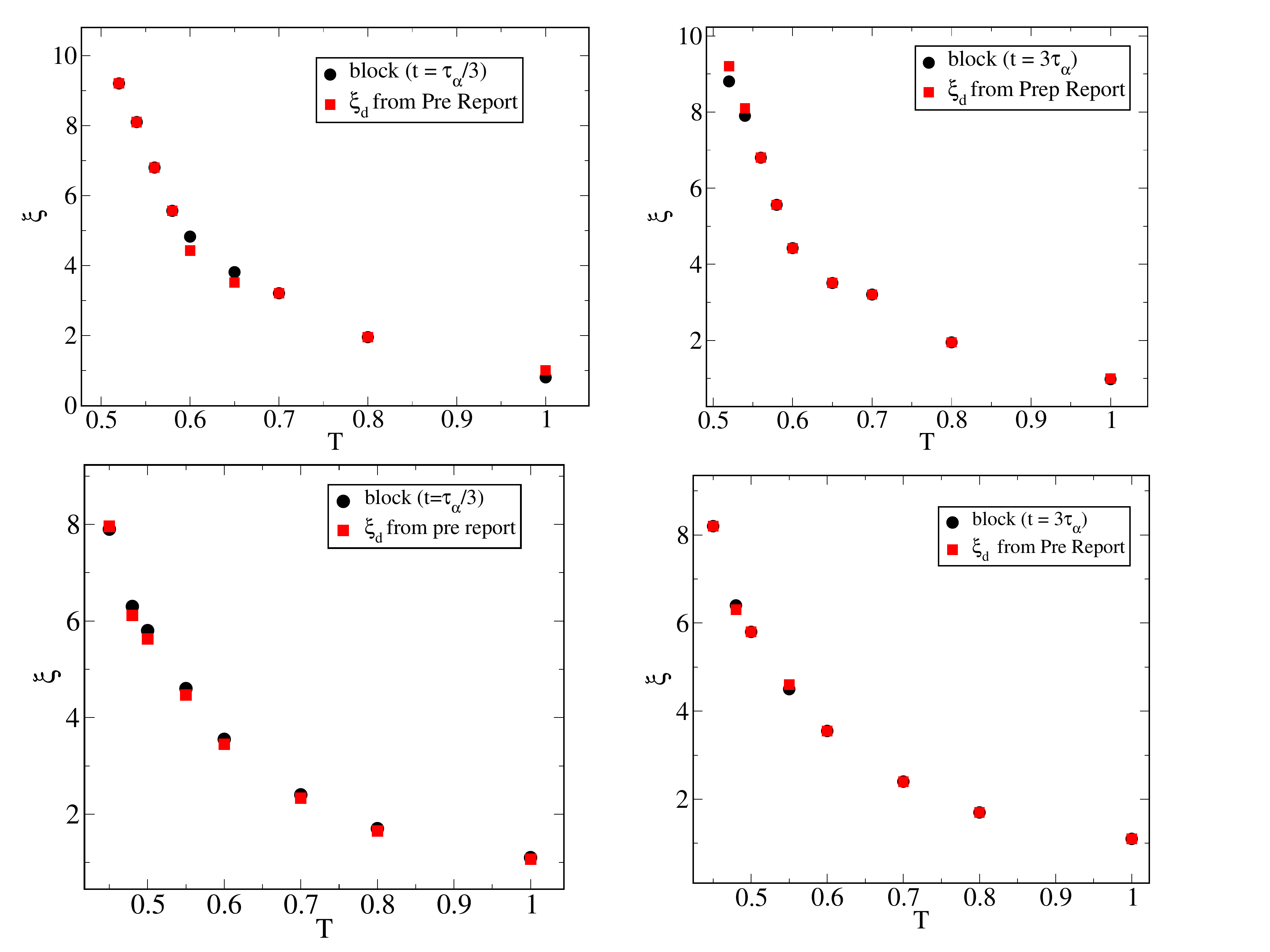}
%\includegraphics[scale=0.405]{com_3dR10_tauby3.pdf}
%\includegraphics[scale=0.415]{com_3dR10_tau3.pdf}
%\vskip +0.5cm
%\begin{center}
%\includegraphics[scale=0.456]{com_3dIPL_tauby3.pdf}
%\includegraphics[scale=0.445]{com_3dIPL_tau3.pdf}
%\end{center}
\caption{\textbf{Top Panel:} Growth of dynamic length scale as a function of temperature for 3dR10 model at time interval $\tau_{\alpha}/3$ and $3\tau_{\alpha}$ obtained from block analysis method and compared with the corresponding quantities which is taken from
Ref.~\cite{PhysRevLett.119.205502}. \textbf{Bottom Panel:} Similar comparison for 3dIPL 
model system. }
  \label{fig:com_3dR10}
\end{center}
\end{figure*}

\vskip +0.2 cm
In Fig.~\ref{fig:scaling3dR10} we plot the block size dependence of 
$\chi_4(L_B,T)$ and its scaling collapse at time scale $\tau_{\alpha}/3$ 
and $3\tau_{\alpha}$ for 3dR10 and 
3dIPL model systems for different temperatures. For all these models 
observed scaling collapse are reasonably good and the calculated 
length scales are very good agreement with the previous reported 
length scales \cite{PhysRevLett.119.205502}. We found 
that temperature dependence of dynamical
length scale $\xi_d$ remains same in between time scale $\tau_{\alpha}/3$ 
to $3\tau_{\alpha}$ as shown in Fig.~\ref{fig:com_3dR10}. This results 
confirm the validity of the results over different model glass forming liquids.  

%\newpage
\section{SCALING FUNCTION AND THE SCALING EXPONENT for 3dR10 and 3dIPL model}

We now examine the power law dependence of dynamical susceptibility 
$\chi_4(T)$ and the correlation length $\xi(T)$ at different time interval 
(e.g. $t = \tau_\alpha/3$ and $3\tau_\alpha$).
The finite size scaling form used as following

\begin{equation}
\chi_4^t(L_B,T) = \chi_0(t) f(\frac{L_B}{\xi_d(T)}) 
\end{equation} 

\begin{figure*}[!htpb]
\begin{center}
\includegraphics[scale=0.50]{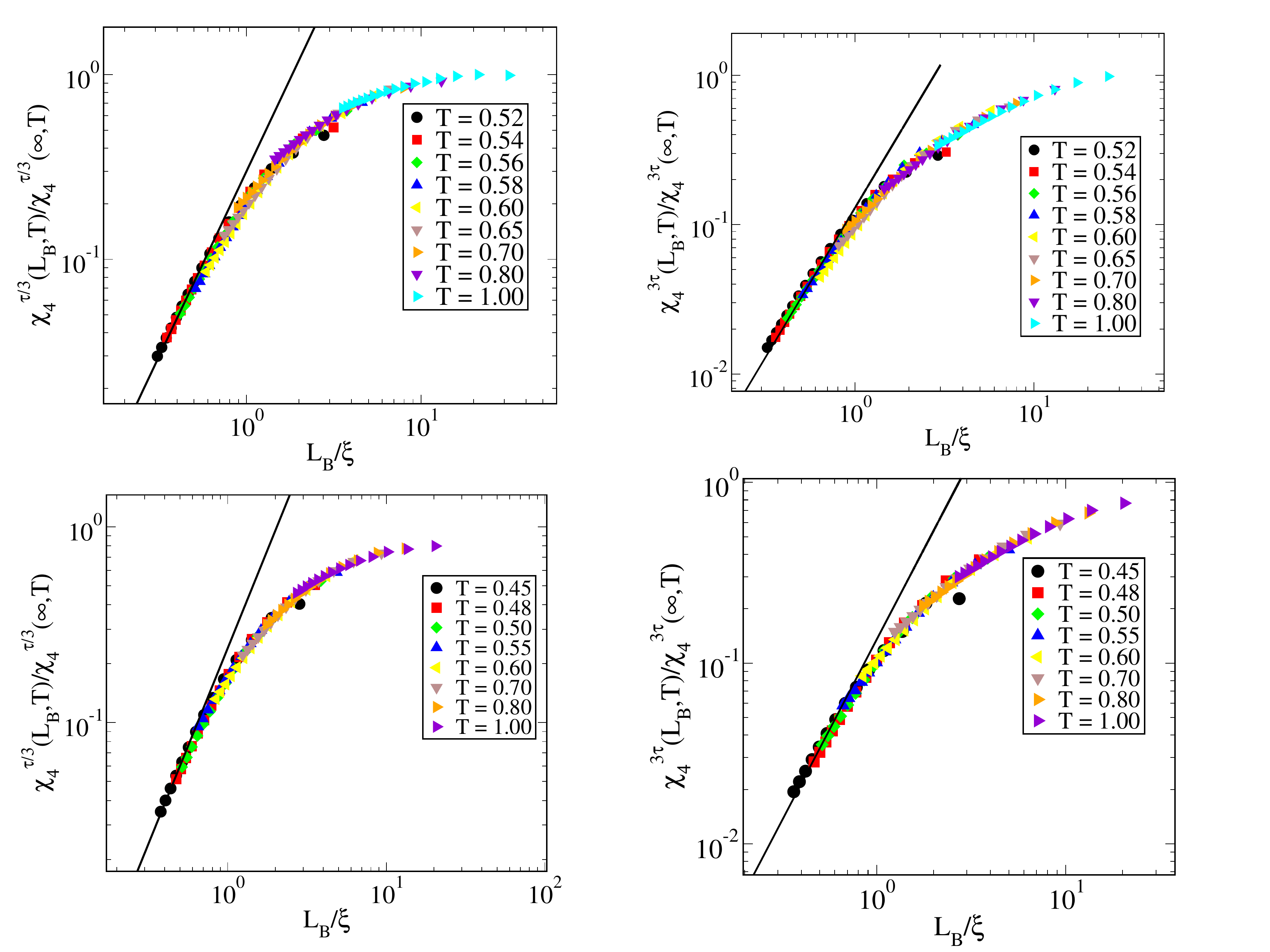}
%\includegraphics[scale=0.38]{x4pvslbyz_tauby3_3dR10.pdf}
%\hspace{0.7mm}
%\includegraphics[scale=0.368]{x4pvslbyz_tau3_3dR10.pdf}
%\vskip +0.3cm
%\includegraphics[scale=0.39]{x4pvslbyz_tauby3_3dIPL.pdf}
%\hspace{0.7mm}
%\includegraphics[scale=0.395]{x4pvslbyz_tau3_3dIPL.pdf}
\caption{\textbf{Top Panel:} Scaling function $f(x) \propto x^{2-\eta}$ for small $x$, we find $\eta=0$ at time interval $t=\tau_{\alpha}/3$ and $t=3\tau_{\alpha}$ for 3dR10 model.  \textbf{Bottom Panel:} Similar analysis done for 3dIPL model.}
  \label{fig:scale_3dR10}
\end{center}
\end{figure*}

Here $\chi_4^t$ denotes the susceptibility at time t. 
$\chi_0(t) = \lim_{L_B \to \infty} \chi_4^t(L_B,T)$. As shown in
\cite{PhysRevLett.119.205502}, we do the similar scaling analysis 
of $\chi_4^t(T)$ to obtain the exponent $(2-\eta)$ at different 
time interval as detailed in the main article for 3dR10 
and 3dIPL model systems. The exponent value is found to be 
$2 - \eta \simeq  2$. This analysis is shown in Fig.~\ref{fig:scale_3dR10}.

\section{Calculation of dynamical length scale from four point structure factor $S_4(q,t)$}
Most of the important information about 
dynamical heterogeneity length scale has been obtained from studies of multi-point 
correlation function and its associated susceptibilities \cite{chandan92}. To extract the 
dynamical length scale many numerical studies consider the four point 
time dependent correlation function defined as:
\begin{equation}
\begin{split}	
g_4(r,t)= \langle \delta \rho(0,0)\delta \rho(0,t)\delta \rho(r,0)\delta \rho(r,t)\rangle -\\ \langle\delta \rho(0,0)\delta \rho(0,t)\rangle \langle\delta \rho(r,0)\delta \rho(r,t)\rangle
\end{split}	
\end{equation}
$\delta \rho(r,t)$ is the deviation of local density $\rho(r,t)$ at position r and time t 
from its average value $\rho_0(r,t)$, $\langle...\rangle$ represents the thermal or time 
average. The fourier 
transform of $g_4(r,t)$ is known as four point time dependent structure factor $S_4(q,t)$ and 
its associated susceptibilities defines as $\lim_{q \to 0}S_4(q,t)\equiv \chi_0(t) $.

To compute the  dynamical heterogeneity length scale we compute four point 
structure factor $S_4(q,t)$ 
\cite{JCP119-14}, which is defined as
\begin{equation}
S_4(q,t) = N [\langle Q(q,t)Q(-q,t)\rangle - \langle Q(q,t)\rangle ^2]
\end{equation}
where 
\begin{equation}
Q(q,t) = \frac{1}{N}\sum_{i=1}^{N}e^{i\vec{q}.\vec{r}_i(0)} 
w(|\vec{r}_i(t)-\vec{r}_i(0)|),
\end{equation}
where $w(x)$ is the same window function defined before. 
$S_4(q,t)$ represents how the dynamics of the particles are spatially 
correlated in the time interval $[0,t]$. The value of $S_4(q,t)$ at 
small $q$ limit ($q \to 0$) increases with decreasing $T$. This 
behaviour is similar  of  that  observed  in static structure factor 
$S(q)$ near liquid-gas critical point where two point density 
fluctuation diverge at small $q$. Near the glass 
transition temperature or density, usual two-point density 
fluctuation does not show any diverging peak. 

\begin{figure*}%[!h]
\begin{center}
\includegraphics[scale=0.8]{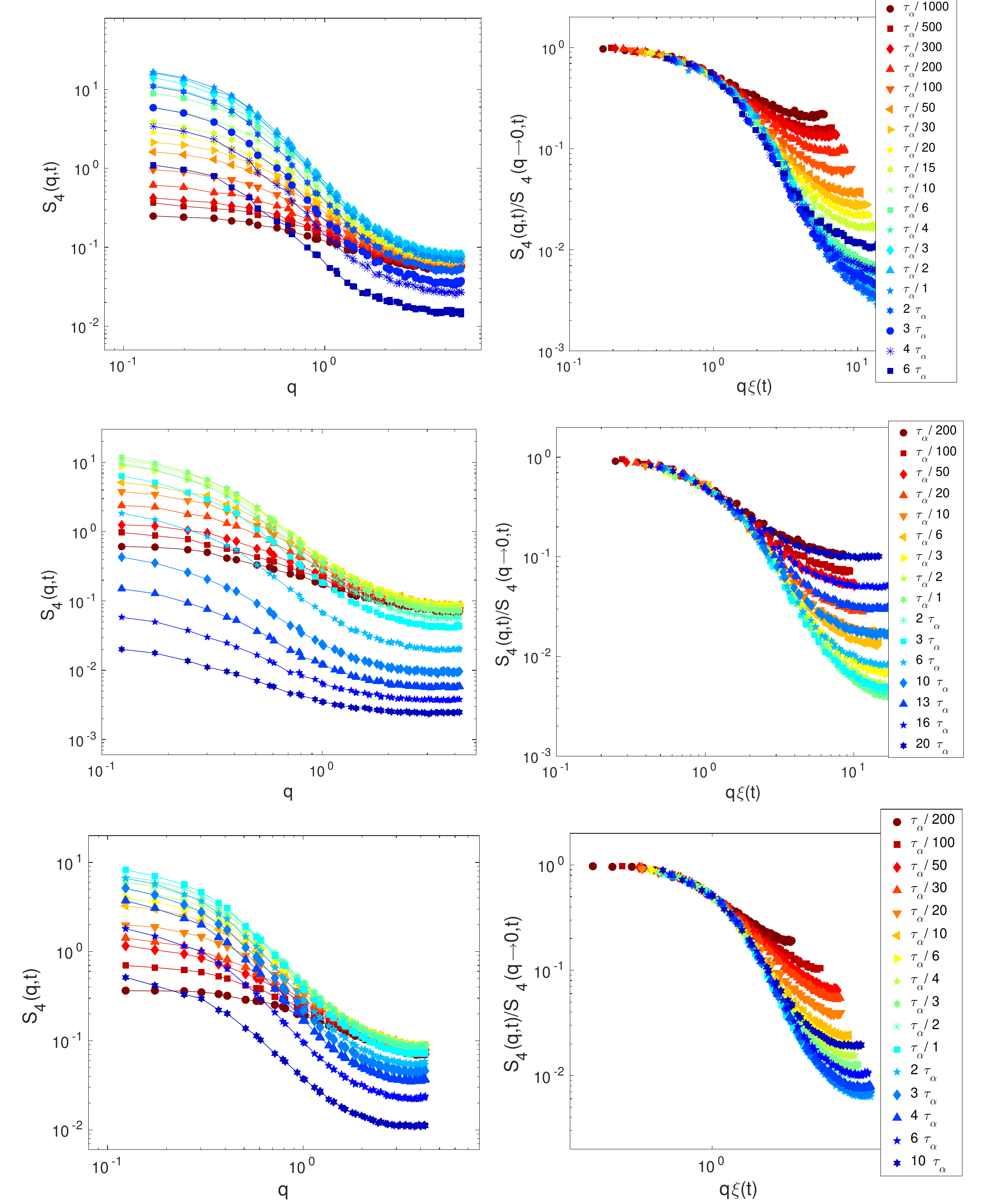}
%\includegraphics[scale=0.34]{3dKA_collapse_difftau.pdf}
%\hspace{-0.7mm}
%\vskip +0.2cm
%\includegraphics[scale=0.34]{3dR10_collapse_difftau.pdf}
%\hspace{-0.7mm}
%\vskip +0.2cm
%\includegraphics[scale=0.38]{3dHP_eps_1_collapse.pdf}
\caption{\textbf{Top Left Panel:} Dynamic structure factor $(S_4(q,t))$ for 3dKA model as a function of wave vector $(q)$ for different time interval at $T = 0.450$. \textbf{Top Right Panel:} Scaling plot $S_4(q,t)/S_4(q \to 0,t)$ vs $q\xi$ for the 108000 particle simulations.  The time interval are shown in the key. The value of $\tau_{\alpha} = 1714.39$. \textbf{Middle Panel:} Similar figure for 3dR10 model at $T = 0.560$. The value of $\tau_{\alpha} = 401.134$. \textbf{Bottom Panel:} Similar figure for 3dHP model at $T = 0.0048$ and $\tau_{\alpha} = 1810.78$.}
  \label{fig:difftaucollapse}
\end{center}
\end{figure*}

In Fig.~\ref{fig:difftaucollapse}, we schematically represents 
the wave vector $(q)$ dependence of $S_4(q,t)$ over different 
time interval. The intensity of heterogeneity increases at low 
$q$ limit. The four-point dynamic correlation length is 
extracted by fitting $S_4(q,t)$ at small $q$ limit with the 
Ornstein-Zernike (OZ) form
\begin{equation}
S_4(q,t) = \frac{S_4(q \to 0,t)}{1+(q\xi)^2}
\end{equation}
%\section{Time dependence of heterogeneity length scale from four point structure factor $S_4(q,t)$}

In left panel of Fig.~\ref{fig:difftaucollapse} we have plotted the wave 
vector ($q$) dependence of four point structure factor $S_4(q,t)$ over 
different time interval in the deep supercooled 
regime. To obtain the length scale we fit $S_4(q,t)$ to the 
Ornstein-Zernike (OZ) form in the range $q\xi \leq 1.0$. Now by using 
the value of $\xi$ obtained from the OZ fit we did scaling plot 
$S_4(q,t)/S_4(q \to 0,t)$ vs $q\xi$ in right panel of 
Fig.~\ref{fig:difftaucollapse} and we find 
a very good scaling collapse for all the model systems. 

\section{Exponent: $\eta$}
\begin{figure}[!h]
\begin{center}
\vskip +0.5cm
\includegraphics[scale=0.4]{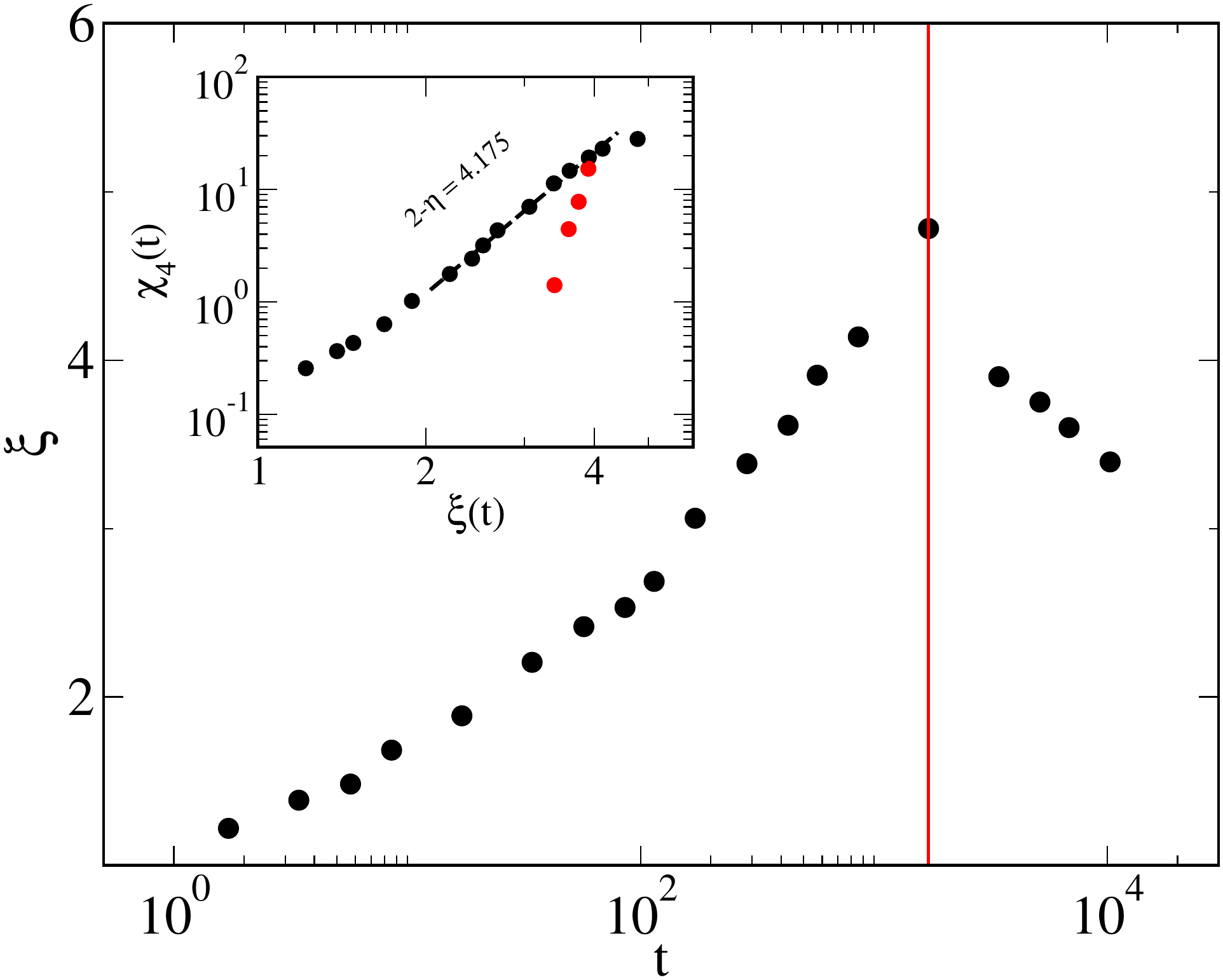}
\includegraphics[scale=0.4]{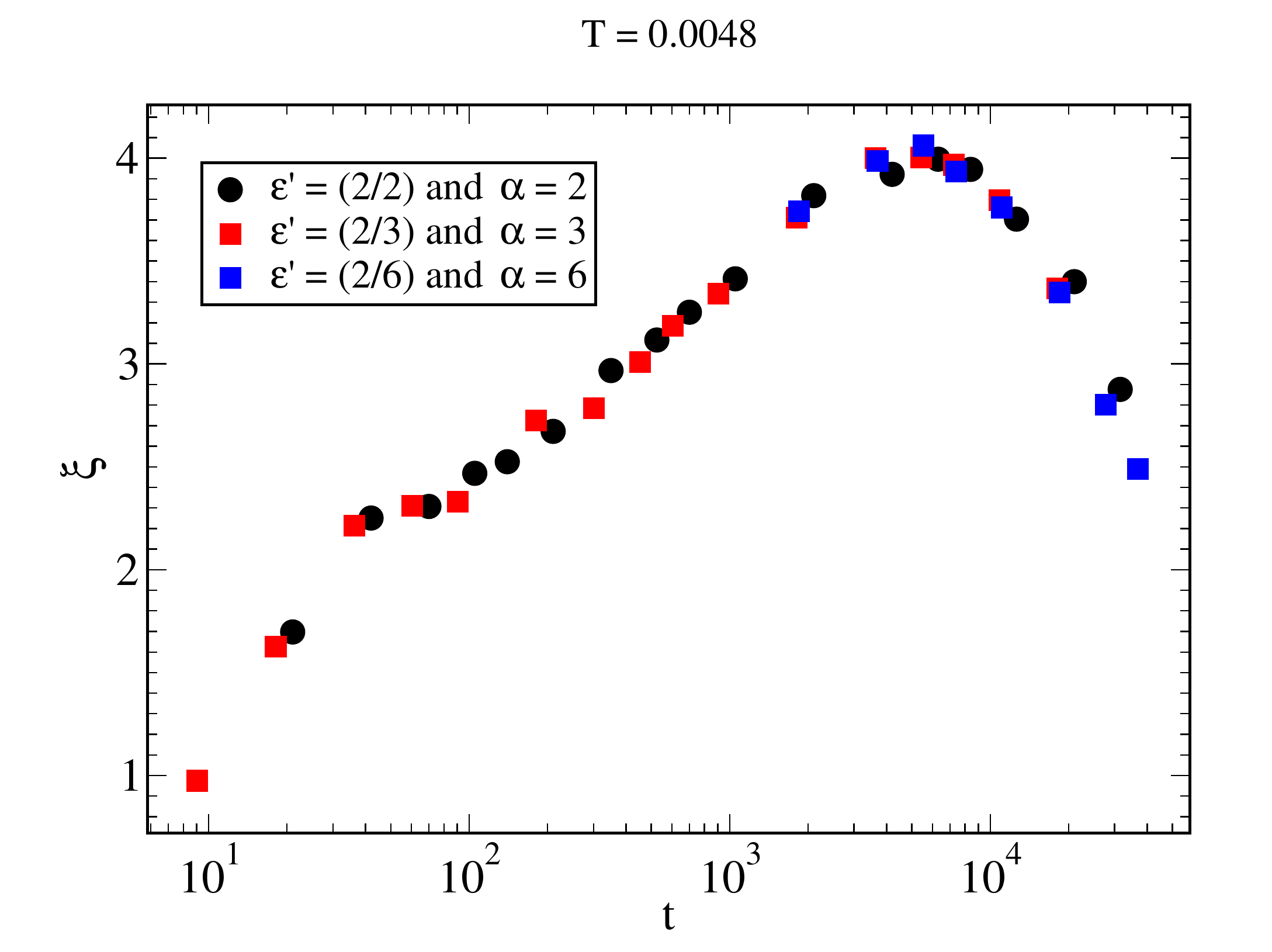}
%\hspace{0.5 cm}
%\includegraphics[scale=0.37]{./chapter3_figs/tvsxi_3dR10.pdf}
\caption{The dynamic length scale vs time for 3dKA model, red line correspond to the $\alpha$ relaxation time. In the inset we show $\chi_4(t)$ vs $\xi_d(t)$.}
  \label{fig:tvsxi}
\end{center}
\end{figure}
We have calculated the exponent associated 
with $\chi_4(t)$ and $\xi_d(t)$, as $\chi_4(t) \sim \xi_d^{2-\eta}$ for
all the models. Results for the 3dR10 and the 3dHP are given in the main article. Here we show the results for the 3dKA model system in
Fig.~\ref{fig:tvsxi}. We obtained the exponent $2-\eta = 4.175$ 
up to $t \sim \tau_{\alpha}$ and a very different behavior for 
$t > \tau_\alpha$ in agreement with the results obtained for 3dR10
model. For the 3dHP model we have shown that $\xi_d(t)$ has a peak
at time $t \simeq 4\tau_\alpha$. On the other hand in 
Ref.\cite{PhysRevE.83.051501} it was shown that $\xi_d(t)$ saturates to a plateau for hard sphere systems. To test whether one will get similar
results if one systematically tunes the interaction potential such that 
one asymptotically approach hard sphere like behaviour by changing the 
parameters of the potential, we did a similar analysis for the same 3dHP 
model but this time by changing the $\alpha$ parameter of the potential.
It is clear that if one takes limit $\alpha \to \infty$, as 
$$\lim_{\alpha \to \infty} \frac{\epsilon}{\alpha} 
\left[1 - \left(\frac{{r}}{\sigma_{\alpha\beta}}\right)\right]^{\alpha},$$ 
then the soft sphere potential will tend towards the hard sphere case. 
Thus we did a similar analysis for two different choices of $\alpha$, 
$\alpha = 3$ and $6$ to see if one obtains hard sphere like results. 
In the bottom panel of Fig.\ref{fig:tvsxi}, we have shown the data once
more to highlight that within the studied range of values of $\alpha$,
the results do not change qualitatively. It is quite possible that 
if one takes much larger values of $\alpha$, then probably one will 
get results similar to the hard sphere, but this is beyond the scope
of the present work and we intend to do this analysis in future. 
%\clearpage
\bibliography{ddcorr_si} 
\bibliographystyle{apsrev4-1}
%bibliographystyle{h-physrev}
%\begin{thebibliography}{10}